\documentclass[journal]{IEEEtran}

\usepackage{amssymb}
\usepackage{graphicx}
\graphicspath{ {images/} }

\usepackage{algorithm}
\usepackage[noend]{algpseudocode}
\usepackage{nccmath}
\usepackage{adjustbox}
\usepackage{comment}
\usepackage{amsmath}
\usepackage{mathtools}
\usepackage{booktabs}
\usepackage{multirow}
\usepackage{tabularx}
\usepackage{amsfonts}
\DeclareMathOperator*{\argmin}{min}

\usepackage{pifont}

\newcommand\tabitem{\makebox[1em][r]{\textbullet~}}
\usepackage{array}
\newcolumntype{P}[1]{>{\raggedright\arraybackslash}p{#1}}
\usepackage{caption} 
\ifCLASSINFOpdf
\else
\fi
\begin{document}
%
\title{Two-stage Deep Denoising with Self-guided Noise Attention for Multimodal Medical Images}
%
%
%

\author{S M A Sharif,~\IEEEmembership{}
        Rizwan Ali Naqvi,~\IEEEmembership{}
        and~Woong-Kee Loh~\IEEEmembership{}
\thanks{This work did not involve human subjects or animals in its research.}
\thanks{S M A Sharif is with Opt-AI Inc., LG Sciencepark, Republic of Korea e-mail: (sharif@opt-ai.kr)}
\thanks{Rizwan Ali Naqvi is with Department of Artificial Intelligence and Robotics, Sejong University, Republic of Korea e-mail: (rizwanali@sejong.ac.kr) }
\thanks{Woong-Kee Loh is with School of Computing, Gachon University, Seongnam, Republic of Korea e-mail: (wkloh2@gachon.ac.kr).}
\thanks{Correspondence: (Rizwan Ali Naqvi; Woong-Kee Loh), Equal contribution: (S M A Sharif; Rizwan Ali Naqvi)}}
\maketitle


\begin{abstract}
Medical image denoising is considered among the most challenging vision tasks. Despite the real-world implications, existing denoising methods have notable drawbacks as they often generate visual artifacts when applied to heterogeneous medical images. This study addresses the limitation of the contemporary denoising methods with an artificial intelligence (AI)-driven two-stage learning strategy. The proposed method learns to estimate the residual noise from the noisy images. Later, it incorporates a novel noise attention mechanism to correlate estimated residual noise with noisy inputs to perform denoising in a course-to-refine manner. This study also proposes to leverage a multi-modal learning strategy to generalize the denoising among medical image modalities and multiple noise patterns for widespread applications. The practicability of the proposed method has been evaluated with dense experiments. The experimental results demonstrated that the proposed method achieved state-of-the-art performance by significantly outperforming the existing medical image denoising methods in quantitative and qualitative comparisons. Overall, it illustrates a performance gain of 7.64 in Peak Signal-to-Noise Ratio (PSNR), 0.1021 in Structural Similarity Index (SSIM), 0.80  in DeltaE ($\Delta E$), 0.1855  in Visual Information Fidelity Pixel-wise (VIFP), and 18.54 in Mean Squared Error (MSE) metrics.
\end{abstract}


%
\IEEEpeerreviewmaketitle

\section{Introduction}
%
%
%
%
\IEEEPARstart{N}oise in medical images refers to the unexpected transformations of pixel values. Such arbitrary modifications can appear in medical images due to numerous inevitable imaging limitations like stochastic capturing conditions, image acquisition process, and respiratory movement \cite{sharif2020learning, geng2021content}. Nonetheless, the noise in medical images can notably degrade the perceptual quality by incorporating errors and artifacts into the results \cite{lei2023ct}. As a result, it impacts the performance of image analysis algorithms, such as segmentation, registration, and classification. Despite having numerous real-world implications, medical image denoising is significantly challenging due to variable noise patterns and imaging characteristics of different imaging modalities \cite{lee2020deep,kulathilake2022review,el2022deep}.
 
\begin{figure*}[!htb]
\centering
\includegraphics[width=0.78\textwidth,keepaspectratio]{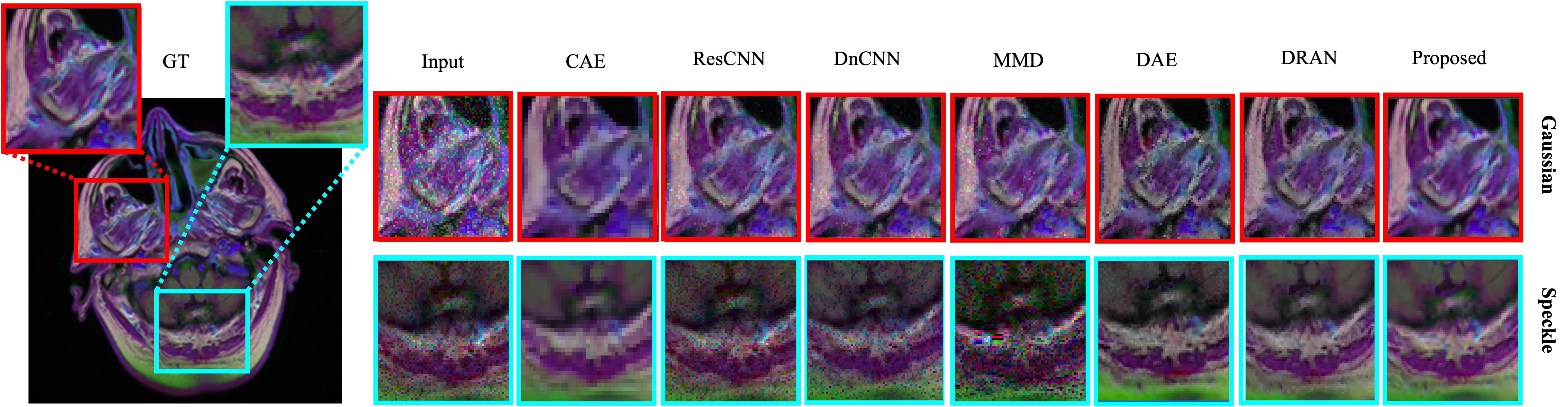}
\caption{ Comparison between deep medical image denoising methods. The existing denoising methods are prone to produce smooth denoising results with visual artifacts. The top row depicts Gaussian denoising; the bottom row shows speckle denoising. In each row, Left to right: GT Image, Noisy Input, CAE \cite{gondara2016medical}, ResCNN \cite{jifara2019medical}, DnCNN \cite{jiang2018denoising}, MMD \cite{el2022deep}, DAE \cite{el2022efficient}, DRAN \cite{sharif2020learning}, and the proposed method. }
\label{intro}
\end{figure*}

Despite the challenges, medical image denoising has drawn significant attention from the vision research community.  In the early days, medical image denoising was tackled with classical image processing techniques like non-local self-similarity (NSS) \cite{wang2006fast}, sparse coding \cite{elad2006image}, and filter-based approaches \cite{arif2011combined,bhonsle2012medical,dabov2007image}. Later, medical image-denoising methods have
advanced beyond the classical approaches by leveraging deep learning with two learning strategies: i) learning denoising as an image-to-image translation by utilizing numerous deep techniques such as encoder-decoder models  \cite{gondara2016medical,el2022efficient,chen2017low,fan2019quadratic}, adversarial learning, \cite{ghahremani2022adversarial, zhou2021mdpet, li2021novel} etc. and ii) learning residual noise from noisy images by stacked convolution models\cite{zhang2017beyond,sharif2020learning,jifara2019medical,el2022deep}.  It is worth noting that the learning-based denoising methods have shown promising consequences compared to traditional approaches. Nevertheless, the performance of these learning-based denoising approaches is also limited to specific image modality and fails in reducing high-intensity noise with variable patterns similar to real-world scenarios (Please see Fig.\ref{intro}).

The existing image-to-image denoising methods are susceptible to producing over-smoothing in complex spatial distribution. Oppositely, the residual denoising strategies illustrate visually disturbing artifacts with the desaturated complex structure. The lack of salient details in the enhanced images can substantially affect the medical image analysis. In addition to the visual artifacts, the existing methods also focus on addressing specific noise types (i.e., gaussian noise \cite{sharif2020learning, jifara2019medical, zhang2017beyond, el2022deep}, speckle noise \cite{guo2011speckle, lee2021multimodal, coupe2009nonlocal}, etc.). These noise-centric denoising approaches fail to address variable noise patterns commonly appearing in medical images. However, a generic denoising strategy with a multipattern cross-modal learning strategy can broaden the range of noise reduction for variable noise patterns of medical images. The limitations of existing methods and the widespread advantages of learning generic denoising motivated this study to introduce a robust learning-based denoising method for medical image modalities. Thus, the proposed method can address the real-world medical image-denoising challenges by leveraging the cross-modal information.

This study proposes a novel two-stage strategy to counter the limitation of medical image denoising by learning denoising in a course to refine manner. Stage I of the proposed method estimates the underlying noise pattern \cite{sharif2020learning}, while stage II utilizes the estimated noise to refine the results with novel self-guided noise attention. The proposed noise attention block strives to perceive the pixel-level correlation between noisy input and estimated noise to enhance the perceptual quality. Apart from the two-stage network, the proposed study also proposes to utilize multi-modal images \cite{guo2019deep,konovalov2023development} to generalize the denoising among numerous image modalities and noise patterns. The feasibility of the proposed method is extensively evaluated in different medical image modalities with dense experiments. The major contribution of the proposed study is as follows:
\begin{itemize}
    \item A two-stage deep network for learning residual noise and leveraging the estimated noise to reconstruct plausible images
    \item A novel noise-attention mechanism to perceive pixel-level attention on estimated noise and given images
    \item Propose to generalize the multi-pattern noise on medical image denoising by leveraging multi-modal images
    \item Conduct dense experiments and achieve state-of-the-art performance on synthesized and real-world medical images.
\end{itemize}

The rest of the paper is organized into four sections. Section \ref{method} details data simulation and learning strategy, Section \ref{results} analyzes the experimental results, and Section \ref{conclusion} concludes this work.

\section{Method}
\label{method}
\begin{figure}[!htb]
\centering
\includegraphics[width=.75\linewidth,keepaspectratio]{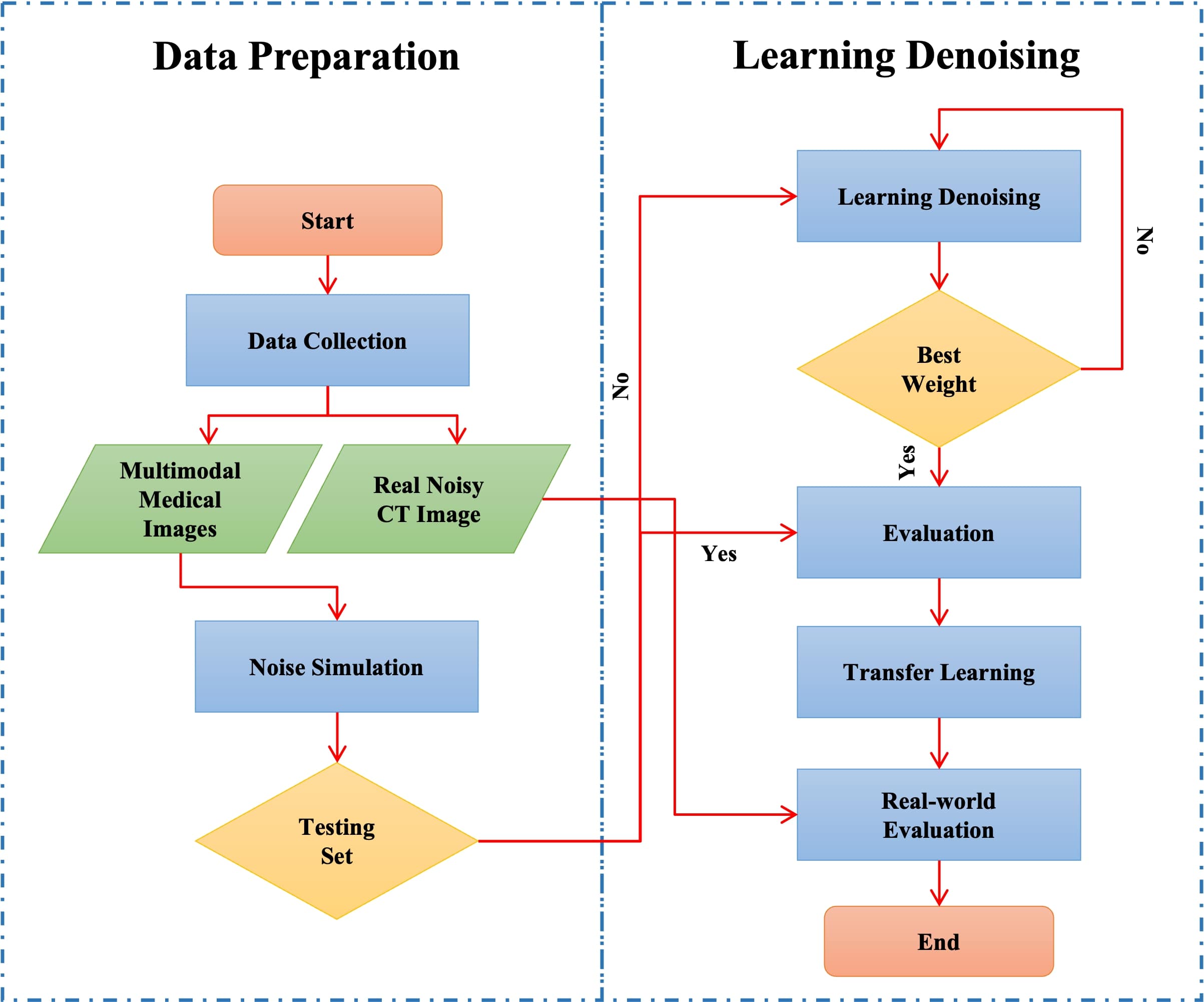}
\caption{ Flowchart of the proposed method. Our proposed method includes data collection and noise simulation strategies. Later, we learned from the collected data and extensively evaluated our method on simulated and real-world noisy medical images. }
\label{flow}
\end{figure}

This study proposes a two-stage novel deep network to learn multi-pattern denoising from multi-modal medical images. This section details the proposed method, including data preparation and evaluation strategies. Fig. \ref{flow} illustrates the flowchart of the proposed method. 

\subsection{Data Preparation}
Data preparation for any medical image task is known to be a challenging task \cite{sharif2022meddeblur, sharif2020learning}. Unfortunately, we could not collect any specialized dataset for studying multi-pattern medical image denoising. Therefore, this study introduces a novel algorithm for synthesizing multi-pattern noise on medical images for pair-wise training. 

\subsubsection{Dataset Collection}
\label{dataInfor}
One of the main objectives of the proposed method is to generalize the medical image denoising among multiple modalities. Therefore, we collected around 30,000 medical images from the open literature before applying the proposed noise-synthesizing algorithm. The collected data samples include multi-modal medical images acquired with different acquisition devices and datasets: i) MRI \cite{buda2019association,lusebrink2021comprehensive}, ii) X-ray \cite{irvin2019chexpert,chouhan2020novel}), iii) CT images \cite{zhao2020covid,clark2013cancer}, iv) Skin lesion images \cite{rezvantalab2018dermatologist}), and v) Microscopy (i.e., protein atlas \cite{uhlen2010towards,spanhol2015dataset}). We used 24,000 random images for training and 1,000 images for validation. The remaining 5,000 samples from collected data (1,000 images from each modality) were used for an extensive evaluation with various noise factors. 

 Apart from the simulated noisy images, we also collected and evaluated our method on real-world noisy CT images (please see section. \ref{real-ct-denoising}).

\subsubsection{Noise Simulation}

Noise in medical imaging can vary depending on the image modality and acquisition conditions \cite{sharif2020learning}. Depending on inclusion,  noise in medical images can be categorized into two types: additive (i.,e Gaussian noise, impulse noise, quantization noise, etc. ) and multiplicative (speckle noise, poison noise, Rician noise, etc.) \cite{gravel2004method}. This study proposes generalizing medical image denoising based on their noise inclusion principle. Therefore, we utilized the most common noise patterns, like Gaussian noise from additive and speckle noise from multiplicative category. It is worth noting that other widely studied noise patterns, such as impulse noise, share many common attributes with the Gaussian and speckle noise. For instance, impulse noise introduces random variations in pixel intensities similar to speckle noise, while its inclusion type shares a similar principle as the Gaussian noise. Overall,  straight learning of Gaussian and speckle denoising helps us illustrate the feasibility of multi-pattern denoising without incorporating every known noise pattern separately. Such a streamlined strategy has enabled us to assess the performance of existing deep models more efficiently and comprehensively.

We generated an noise (${n}_s$) on a given noise-free image ${c}$ as:

\begin{equation}
 {n}_s \sim \mathcal{N}({c}|\mu, \sigma^2)
\end{equation}
$\mu$ and $\sigma^2$ represent a noise distribution's mean and variance ($\mathcal{N}$). 
We generated the noise pattern for learning denoising using the generated noise distribution.

Simulation of Gaussian Noise:
Gaussian noise is known as the most common form of additive noise \cite{sharif2020learning}. Typically, it  derived as: 
\begin{equation}
\label{eqgauss}
 {I_N} = {I_C} + {n}_s
\end{equation}
Here, ${I_C}$ represents the given clean image.

Simulation of Speckle noise:
Speckle noise is the most common form of multiplicative noise observed in medical images. Typically, it is caused by the interaction of light or sound waves with small particles or irregularities in the imaging medium \cite{guo2011speckle}. The speckle noise in medical images can be derived as follows:

\begin{equation}
\label{eqspeckle}
 {I_N} = {I_C} + {I_C} \times {n}_s
\end{equation}

\begin{figure}[!htb]
\centering
\includegraphics[width=.8\linewidth,keepaspectratio]{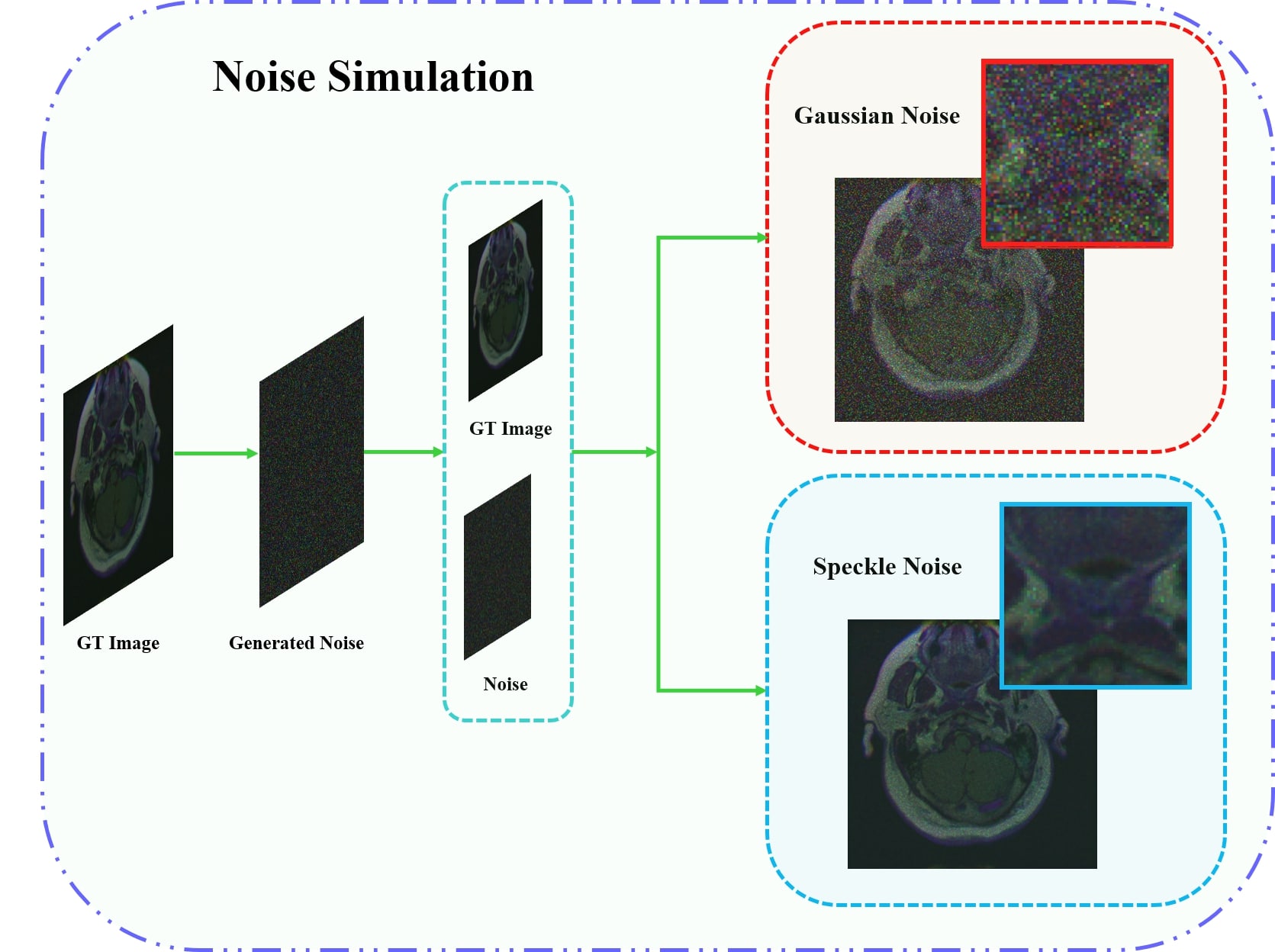}
\caption{ Overview of noisy image generation. The proposed simulation method generates noisy-clean image pairs for training and evaluation. }
\label{simulation}
\end{figure}

Based on eqn. \ref{eqgauss} and \ref{eqspeckle}, this study proposed to generate the random noise patterns for learning multi-pattern denoising, as shown in Algo. \ref{dataSyth}. The simulation algorithm generates an arbitrary standard deviation of noise distribution between 0 and 75. This standard deviation has been utilized to generate multi-pattern noise in the given images. Also, the proposed algorithm ensures an even portion of noise patterns in the learning dataset. Notably, such an approach ensures that the generated noise varies in intensity, providing a realistic representation of the noise commonly found in real-world scenarios. Fig. \ref{simulation} illustrates the sample image-pairs generated by the proposed noise simulation algorithm.

\begin{algorithm}

\caption{Noise simulation in medical image simulation}\label{alg:euclid}
\begin{algorithmic}[24]

\State $Mean$ = 0 - mean for generating random noise

\Procedure{SIMULATION}{$I_G, Mean$}
\State $Std\gets random(0, 75)$ - standard deviation for generating random noise

\State $Noise \gets normalDistribution(Mean, Std)$ - generating random noise

\State $noisePattern\gets random(0, 10)$

\If{$noisePattern \% 2 == 0$}
\State $I_N\gets I_G + Noise$ - generating Gaussian noise
\Else{}
   \State $I_N\gets I_G + (I_G \times  Noise)$ - generating speckle noise
\EndIf

\State \textbf{return} $I_N$
\EndProcedure
\end{algorithmic}
\label{dataSyth}
\end{algorithm}

\subsection{Learning Denoising}

The medical image denoising is typically handled with the single-stage network. Unlike the previous medical image-denoising methods, this study proposes to leverage a two-stage approach to learn medical image-denoising in a course to refine manner. The proposed method aims to denoise a given noisy medical image as $\mathrm{D}: I_N \to  I_C$. Here, the learning function ($\mathrm{D}$) learns to generate a clean image ($I_C$) from a given noisy image ($I_N$). 

\begin{figure*}[!htb]
\centering
\includegraphics[width=.75\textwidth,keepaspectratio]{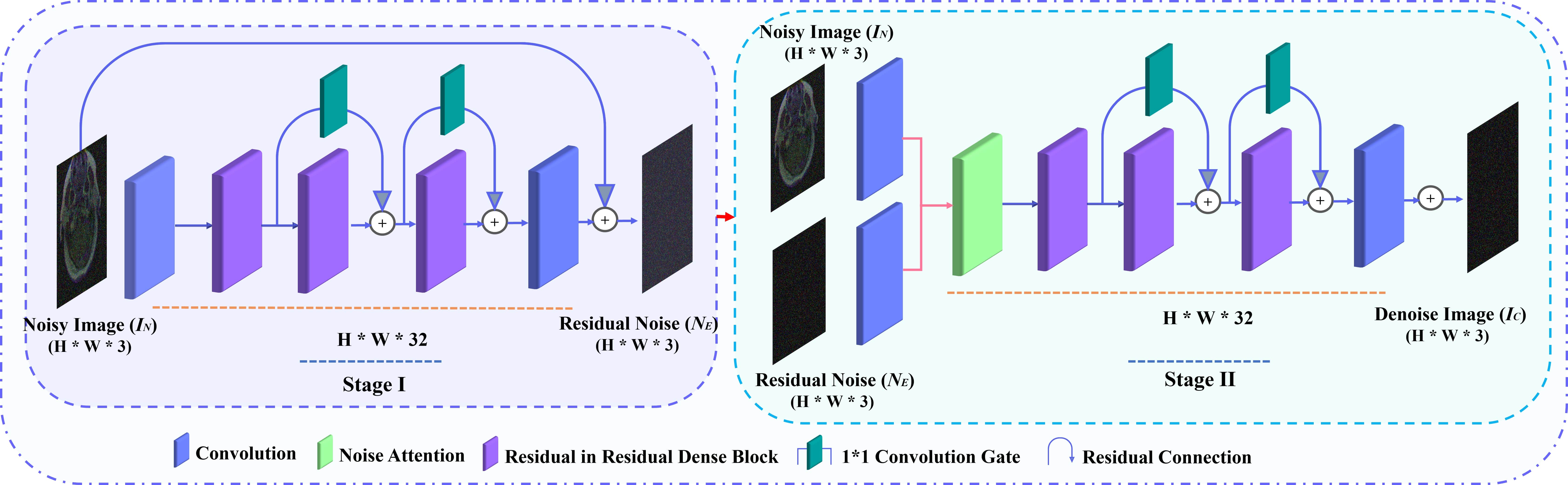}
\caption{ Overview of the proposed two-stage network. Stage I of the proposed method learns to estimate the residual noise, and stage II leverages the estimated noise in refining the denoise outputs. }
\label{net}
\end{figure*}

As Fig. \ref{net} illustrates, the learning function ($\mathrm{D}$) comprises two learning stages with two deep networks. The stages of the proposed method aim to perform as follows:

\begin{itemize}
    \item \textbf{Stage I:}  Stage I of the proposed method takes a noisy medical image ($\mathrm{I_N}$) and estimates the underlying noise ($\mathrm{N_E}$). It comprises a single deep network denoted as a noise estimation network (NEN). 
    
    \item \textbf{Stage II:} Stage II of the proposed method aims to reconstruct the clean image ($I_C$) by leveraging the estimated noise ($\mathrm{N_E}$) and the noisy image ($\mathrm{I_N}$). Stage II of the proposed method utilizes a novel self-guided noise attention block to estimate the correlation between the noise patterns of the given inputs.  It has been denoted as a reconstruction network (RN) in the later section.
\end{itemize}

\subsubsection{Noise Estimation Network (NEN)}

The noisy medical image is typically derived as follows:

\begin{equation}
{I_N = I_C  \oplus N}
\label{resDerive}
\end{equation}

Here, $\oplus$ represents the types of noise inclusion (i.e., additive or multiplicative), ${I_N}$ represents the noisy image, $N$ is the underlying noise, and ${I_C}$ is the underlying clean image. The proposed NEN aims to estimate the underlying residual noise ($N_E$) from a given noisy medical image ${I_N}$  through the mapping function ${\mathrm{F_E}: I_N \to N_E}$.  Mapping function ($\mathrm{F_E}$) learns the residual noise pattern ($N_E$) as $N_E \in [0,1]^{H \times W \times 3}$. $H$ and $W$ represent the height and width of the output ($N_E$).

To estimate the underlying noise pattern ($N_E$), this study leverages consecutive residuals in residual dense blocks (RRDB) \cite{wang2018esrgan}. Notably, the RRDB blocks are well known for tackling the gradient vanishing problems. It aims to help the proposed method achieve a deeper architecture by retaining the flow of gradient information. Overall, the RRDB has been pursued as follows:

\begin{equation}
R_{X} = H_{R,d}((H_{R,d-1}(R_0)))
\end{equation}

where $H_R$ denotes the residual dense blocks (RDB) \cite{zhang2018residual}, and $d$ denotes the operations of the $d^{th}$ RDB, which again can be  formulated as follows:
\begin{equation}
R_{d,c} =\tau(H_{d,c}([R_{d-1},R_{d,1},...,R_{c-1}]))
\end{equation}

Here $[R_{d-1},R_{d,1},...,R_{c-1}]$ refers to the concatenated
feature maps obtained through the $d=5$ number of convolutional layers, $c=8$ denotes the growrate.  All convolution layers incorporate the kernel = $3 \times 3$, stride = 1, padding = 1, and are activated with the LeakyReLU function. Apart from the RRDB, we leverage the noise gate with $1 \times 1$ convolution operation to utilize short-distance residual feature propagation between consecutive RRDB blocks.

Estimating residual noise for denoising has illustrated significant momentum in recent studies. However, denoising based on only residual noises can demonstrate a few drawbacks. For instance, an underestimated residual noise can lead to final output in producing the over-smoothing regions. Similarly, overestimated residual noise is well known for producing over-sharpen final images. Which again can introduce the visual artifacts. This study proposes introducing a secondary learnable network to address the limitation of residual noise estimation.

\subsubsection{Reconstruction Network (RN)}

The proposed RN aims to recover clean medical images ($I_C$) by leveraging the estimated noise ($N_E$) along with the noisy input image  ($I_N$) as  ${\mathrm{F_R}: {I_N, N_E} \to I_C}$. The mapping function ($\mathrm{F_R}$) learns to generate a clean medical image ($I_C$) as $I_C \in [0,1]^{H \times W \times 3}$. $H$ and $W$ represent the height of the reconstructed clean image. 

Overall, it shares a network architecture similar to NEN. However, unlike the NEN, it takes two inputs: estimated noise ($N_E$) and noisy image ($I_N$). It applies a standard $3 \times 3$ convolution operation to feed the mapped features into the proposed self-guided noise attention block. Later, the attention map of the noise attention block (NOB) propagated into consecutive RRDBs to generate noise-free images \cite{khowaja2023face}. 

\begin{figure}[!htb]
\centering
\includegraphics[width=.8\linewidth,keepaspectratio]{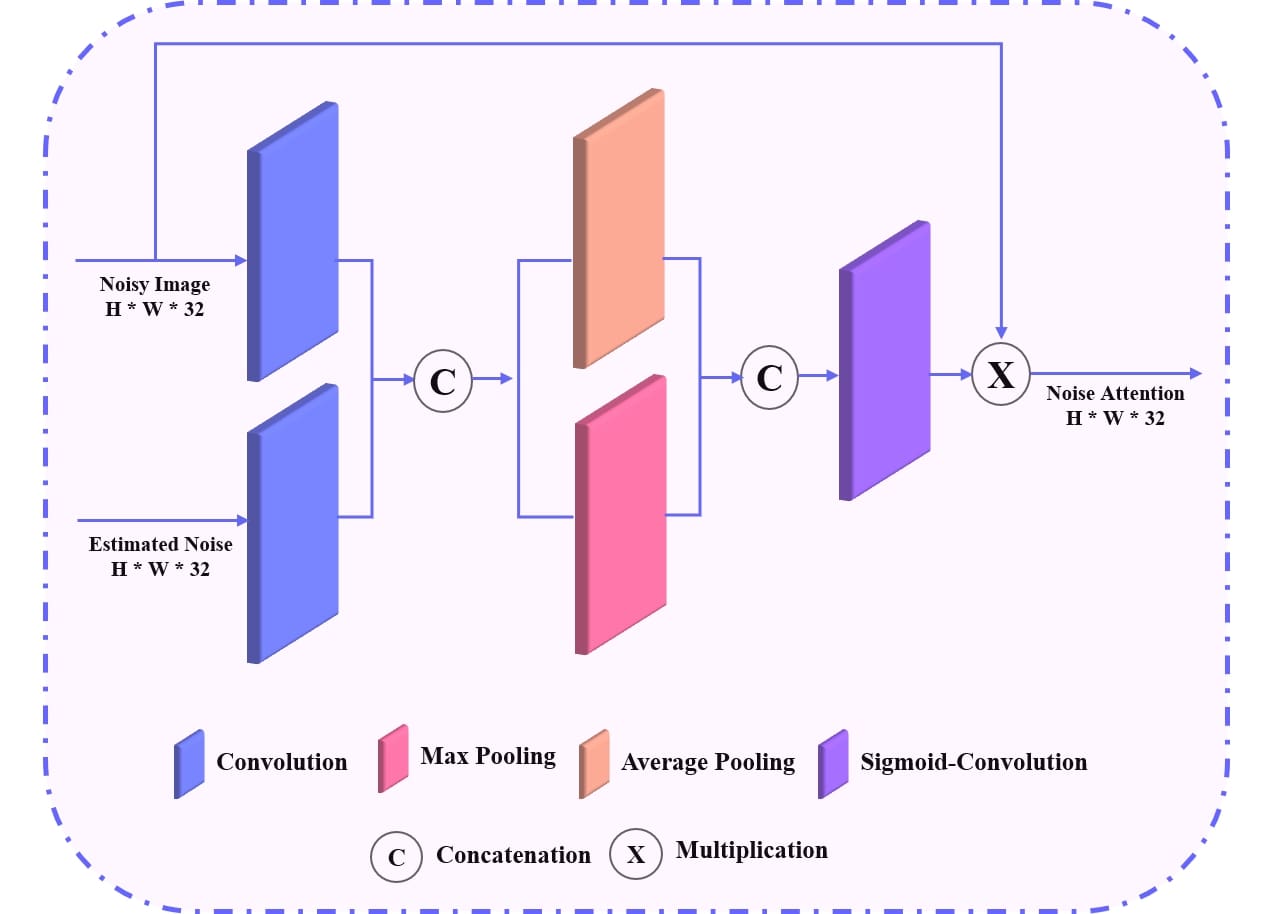}
\caption{ The architecture of the proposed noise attention block. It estimates the pixel-level attention over noisy input with a residual noise pattern.}
\label{attention}
\end{figure}

\textbf{Noise Attention.} The proposed novel self-guided NOB intends to accelerate denoising performance. The NOB leverages the estimated noise features and perceives pixel-level attention on the high-intensity noise region as shown in Fig. \ref{attention}. The noise correlation has been perceived as follows:

\begin{equation}
{X_C} = {C}({N_T}) \oplus {C}({I_T})
\end{equation}

Where, $C(\dot)$ represents the convolution operation with kernel=3, stride=1, filter=32, $\oplus$ = concatenation, ${N_T}$ = estimated feature map of residual noise, and ${I_T}$ = convolutional feature of noisy input. 

The pixel level spatial attention \cite{woo2018cbam,sharif2021sagan} over concatenated feature map ${X_C}$ has been estimated as:

\begin{equation}
{A_C} = {Z_S}(
[{Z_{A}(X)};{Z_M(X)}]
)
\end{equation}

Where ${Z_{A}}$ represents the convolution operation with sigmoid activation. Also, ${Z_{A}}$ and ${Z_{M}}$ present the average pooling and max pooling, which generates two 2D feature maps as ${X_{A}} \in \mathbb{R}^{1 \times H \times W}$, ${X_{M}} \in \mathbb{R}^{1 \times H \times W}$. 

The final 32 channels of self-guided noise attention have been perceived as below:

\begin{equation}
{S_A} = {A_C} \times  {I_T} 
\end{equation}

\subsubsection{Learning objectives}

The proposed NEN and RN are optimized with separate objective functions. The objective of NEN can be derived as follows:
 
\begin{equation}
 W_E^\ast = \arg{\argmin_W}\frac{1}{P}\sum_{t=1}^{P}\mathcal{L}_{\mathit{E}}({N_E}^t), {N_G}^t)
\end{equation}

Here, $\mathcal{L}_\mathit{E}$ has been set as mean-square-error loss as :

\begin{equation}
 \mathcal{L}_\mathit{E} = \parallel {N_G} - {N_E} \parallel_1^2
\end{equation}

Similarly, the objective function of RN  aims to minimize the error as follows:

\begin{equation}
 W_R^\ast = \arg{\argmin_W}\frac{1}{P}\sum_{t=1}^{P}\mathcal{L}_{\mathit{R}}({I_N}^t), {I_G}^t)
 \label{fLoss}
\end{equation}

Here, $\mathcal{L}_\mathit{R}$ has been set as L1-loss:

\begin{equation}
 \mathcal{L}_\mathit{R} = \parallel {I_G} -{I_C} \parallel_1
\end{equation}

It is worth noting  L1 and L2 distances are commonly used as pixel-wise loss functions. However,  L2-loss is directly related to the PSNR and tends to produce smoother images \cite{sharif2023darkdeblur}. Therefore, an L1 objective function has been considered as a reconstruction loss for the RN throughout this study.

\subsubsection{Training Details}

The proposed two-stage multi-pattern denoising is implemented with the PyTorch framework \cite{pytorch}. The NEN and RN were optimized with an  Adam optimizer \cite{kingma2014adam}. The hyperparameters for the optimizer were tuned as $\beta_1 = 0.9$, $\beta_2 = 0.99$, and learning rate = 5e-4.  We trained both models for 100 epochs with a constant batch size of 24. During the training process, we generated random patterns with random noise standard deviation. Please see Algorithm. \ref{dataSyth} for noise simulation and learning procedure of the proposed method. Notably, such random noise generation techniques help the proposed method in addressing the overfitting until model convergence \cite{sharif2019deep}. It took around 24 hours to converge our model. We conducted our experiments on a machine comprised of an AMD Ryzen 3200G central processing unit (CPU) clocked at 3.6 GHz, a random-access memory of 16 GB, and An Nvidia Geforce GTX 3060 (12GB) graphical processing unit (GPU).

\section{Experiments and results}
\label{results}
The feasibility of the proposed method has been verified with dense experiments. We studied the performance of state-of-the-art medical image denoising methods on five image modalities (i.e., MRI, X-ray, Microscopy, Skin Lesion, and CT) images by incorporating Gaussian and speckle noise with distinct noise levels (i.e., 20, 25, 50, and 75) to each image sample. Further, we assessed the comparing method with qualitative comparison and summarised quantatitive performance \cite{roy2023multiclass,wang2004image} with the following evaluation metrics: PSNR, SSIM,  DeltaE, VIFP, and MSE. Later, we verified the feasibility of the novel components with ablation experiments.

\subsection{Comparison with State-of-the-art}
We compared the performance of multi-pattern medical image denoising with state-of-the-art residual denoising methods (i.e., ResCNN \cite{jifara2019medical}, DnCNN \cite{jiang2018denoising},  MMD \cite{el2022deep},
MID-DRAN \cite{sharif2020learning}) and image-to-image translation medical image denoising method (i.e., CAE \cite{gondara2016medical}, DAE \cite{el2022efficient}).  All denoising methods have been trained and tested under the same data samples with suggested hyperparameters. 

\subsubsection{MRI}
MRI is one of the most widely used medical imaging techniques that uses a powerful magnetic field and radio waves to create detailed images of the inside of the body \cite{nguyen2012denoising}. MRI is used to diagnose numerous medical conditions, including cancer, heart disease, neurological disorders, and musculoskeletal injuries. Table \ref{MRITab} illustrates the performance comparison between state-of-the-art deep denoising models and the proposed method for MRI denoising.  The proposed two-stage network outperforms the existing denoising method for speckle and Gaussian denoising. It cumulatively outperforms the deep denoising methods by 11.04 in PSNR, 0.0358 in SSIM, 1.67 in $\Delta E$, 0.2843 in VIFP, and 27.62 in MSE metric for speckle denoising. It also illustrates a significant jump in Gaussian denoising by 8.2725 in PSNR, 0.0966 in SSIM, 1.67 in $\Delta E$, 0.2297 in VIFP, and 33.34 in MSE.

\begin{figure*}[!htb]
\centering
\includegraphics[width=0.65\textwidth,keepaspectratio]{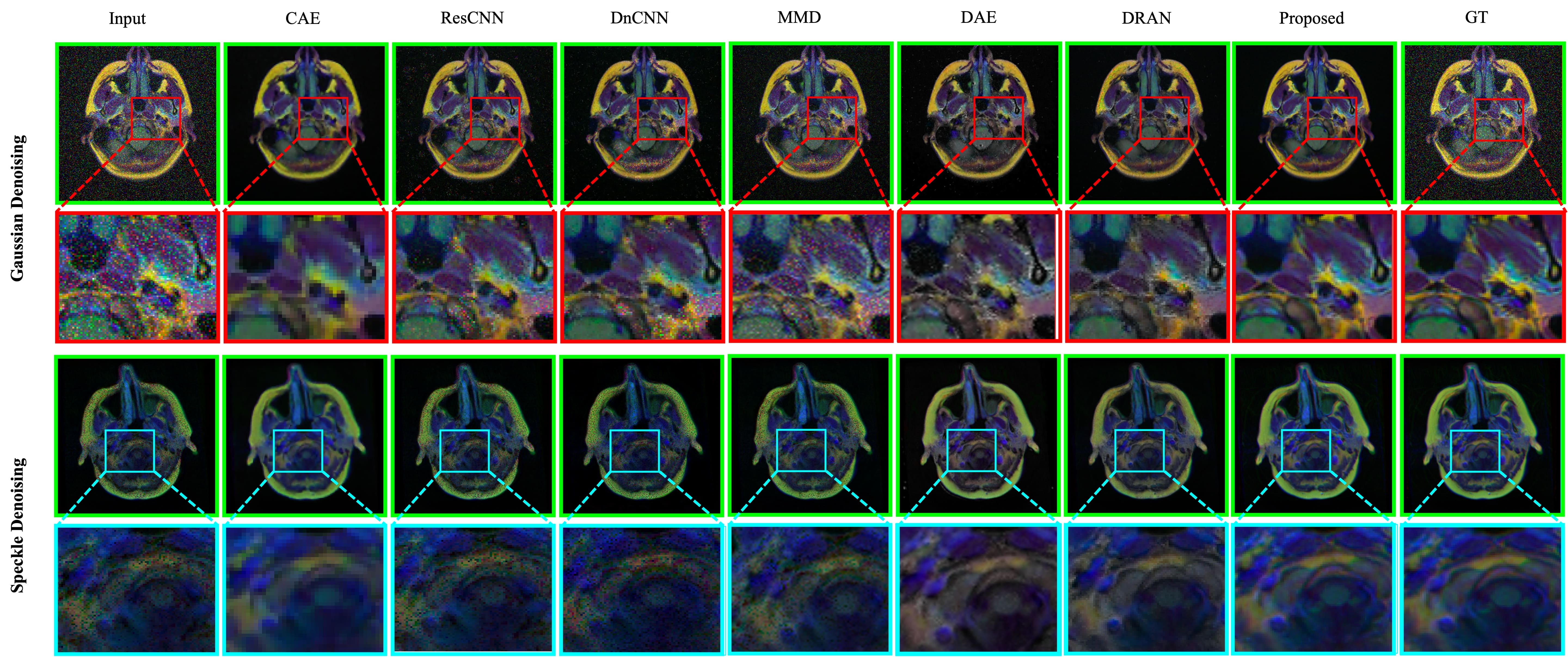}
\caption{Qualitative comparison for MRI denoising. The proposed method can produce sharper images without incorporating visual artifacts. The top example illustrates Gaussian denoising; the bottom shows speckle denoising. In each row, Left to right: Noisy input, CAE \cite{gondara2016medical}, ResCNN \cite{jifara2019medical}, DnCNN \cite{jiang2018denoising}, MMD \cite{el2022deep}, DAE \cite{el2022efficient}, DRAN \cite{sharif2020learning} \cite{sharif2020learning}, proposed method, and ground truth image. }
\label{mri}
\end{figure*}

Additionally, we compared the deep methods visual comparison, as shown in Fig. \ref{mri}. It is visible that the proposed method can produce visually plausible images without illustrating noticeable visual artifacts.

\begin{table*}[!htb]
\centering
\scalebox{.52}{\begin{tabular}{lclllllllllllllll}
\hline
\multirow{2}{*}{\textbf{Model}}              & \multirow{2}{*}{\textbf{$\sigma$}}  & \multicolumn{5}{l}{\textbf{Gaussian}}                                                                                                                                                   & \multicolumn{5}{l}{\textbf{Speckle}}                                                                                                                                                    & \multicolumn{5}{l}{\textbf{Average}}                                                                                                                                                    \\ \cmidrule(){3-17}
                                   &                           & \textbf{PSNR}  & \textbf{SSIM}   & \textbf{$\Delta E$} & \textbf{VIFP}   & \textbf{MSE}  & \textbf{PSNR}  & \textbf{SSIM}   & \textbf{$\Delta E$} & \textbf{VIFP}   & \textbf{MSE}  & \textbf{PSNR}  & \textbf{SSIM}   & \textbf{$\Delta E$} & \textbf{VIFP}   & \textbf{MSE}  \\ \hline
CAE                                & \multirow{7}{*}{10}       & 33.93                           & 0.9375                           & 2.42                                 & 0.4071                           & 37.63                          & 33.82                           & 0.9381                           & 2.40                                 & 0.4094                           & 38.86                          & 33.87                           & 0.9378                           & 2.41                                 & 0.4083                           & 38.24                          \\
ResCNN                             &                           & 26.68                           & 0.7517                           & 5.07                                 & 0.2261                           & 173.40                          & 24.48                           & 0.6878                           & 6.28                                 & 0.1615                           & 298.81                         & 25.58                           & 0.7198                           & 5.68                                 & 0.1938                           & 236.11                         \\
DnCNN                              &                           & 26.53                           & 0.7131                           & 5.71                                 & 0.2197                           & 180.03                         & 25.13                           & 0.6934                           & 6.05                                 & 0.1708                           & 256.01                         & 25.83                           & 0.7033                           & 5.88                                 & 0.1953                           & 218.02                         \\
MMD                &                            & 28.40                           & 0.8395                           & 3.99                                 & 0.5339                           & 191.15                         & 28.40                           & 0.8395                           & 3.99                                 & 0.5339                           & 191.15                         & 28.40                           & 0.8395                           & 3.99                                 & 0.5339                           & 191.15                         \\
DAE                &                             & 24.88                           & 0.6848                           & 6.33                                 & 0.1899                           & 277.58                         & 23.32                           & 0.6280                           & 7.34                                 & 0.1471                           & 391.25                         & 24.10                           & 0.6564                           & 6.83                                 & 0.1685                           & 334.41                         \\
DRAN                               &                           & 35.15                           & 0.9442                           & 2.52                                 & 0.5164                           & 28.23                          & 35.49                           & 0.9613                           & 2.18                                 & 0.5878                           & 30.00                          & 35.32                           & 0.9528                           & 2.35                                 & 0.5521                           & 29.12                          \\
\textbf{Proposed} &                           & \textbf{43.00} & \textbf{0.9851} & \textbf{0.94}       & \textbf{0.7208} & \textbf{3.92} & \textbf{50.61} & \textbf{0.9977} & \textbf{0.34}       & \textbf{0.9135} & \textbf{0.8800} & \textbf{46.81} & \textbf{0.9914} & \textbf{0.64}       & \textbf{0.8172} & \textbf{2.40} \\ \hline
CAE                                & \multirow{7}{*}{25}       & 33.72                           & 0.9318                           & 2.51                                 & 0.3915                           & 38.48                          & 33.93                           & 0.9386                           & 2.38                                 & 0.41                             & 37.87                          & 33.83                           & 0.9352                           & 2.45                                 & 0.4008                           & 38.18                          \\
ResCNN                             &                           & 28.84                           & 0.7704                           & 4.00                                 & 0.3069                           & 91.43                          & 24.57                           & 0.6896                           & 6.26                                 & 0.1649                           & 290.96                         & 26.70                           & 0.73                             & 5.13                                 & 0.2359                           & 191.20                         \\
DnCNN                              &                           & 27.82                           & 0.7689                           & 4.46                                 & 0.2694                           & 118.41                         & 25.44                           & 0.7079                           & 5.83                                 & 0.1823                           & 237.40                         & 26.63                           & 0.7384                           & 5.15                                 & 0.2259                           & 177.91                         \\
MMD                &                            & 28.43                           & 0.8379                           & 3.98                                 & 0.5309                           & 188.05                         & 28.43                           & 0.8379                           & 3.98                                 & 0.5309                           & 188.05                         & 28.43                           & 0.8379                           & 3.98                                 & 0.5309                           & 188.05                         \\
DAE                &                             & 27.98                           & 0.7709                           & 4.56                                 & 0.2908                           & 134.03                         & 23.53                           & 0.6384                           & 7.11                                 & 0.1551                           & 370.02                         & 25.75                           & 0.7047                           & 5.84                                 & 0.2229                           & 252.02                         \\
DRAN                               &                           & 34.38                           & 0.9270                           & 2.60                                 & 0.4799                           & 30.96                          & 35.58                           & 0.9604                           & 2.21                                 & 0.5755                           & 29.18                          & 34.98                           & 0.9437                           & 2.40                                 & 0.5277                           & 30.07                          \\
\textbf{Proposed} &                           & \textbf{41.77} & \textbf{0.9830}  & \textbf{1.03}       & \textbf{0.6865} & \textbf{5.54} & \textbf{47.42} & \textbf{0.9958} & \textbf{0.50}       & \textbf{0.8677} & \textbf{1.73} & \textbf{44.60} & \textbf{0.9894} & \textbf{0.77}       & \textbf{0.7771} & \textbf{3.64} \\ \hline
CAE                                & \multirow{7}{*}{50}       & 33.28                           & 0.9211                           & 2.67                                 & 0.3653                           & 41.62                          & 33.99                           & 0.9387                           & 2.37                                 & 0.4091                           & 37.41                          & 33.64                           & 0.9299                           & 2.52                                 & 0.3872                           & 39.51                          \\
ResCNN                             &                           & 26.75                           & 0.6155                           & 5.14                                 & 0.3157                           & 138.83                         & 24.97                           & 0.7032                           & 5.97                                 & 0.1778                           & 260.36                         & 25.86                           & 0.6594                           & 5.56                                 & 0.2468                           & 199.60                         \\
DnCNN                              &                           & 26.83                           & 0.7171                           & 4.01                                 & 0.3145                           & 136.33                         & 26.23                           & 0.7438                           & 5.24                                 & 0.2097                           & 194.76                         & 26.53                           & 0.7305                           & 4.63                                 & 0.2621                           & 165.54                         \\
MMD                &                            & 28.45                           & 0.8343                           & 3.98                                 & 0.5215                           & 183.68                         & 28.45                           & 0.8343                           & 3.98                                 & 0.5215                           & 183.68                         & 28.45                           & 0.8343                           & 3.98                                 & 0.5215                           & 183.68                         \\
DAE                &                             & 26.33                           & 0.6137                           & 5.52                                 & 0.3097                           & 156.64                         & 23.90                           & 0.6550                           & 6.73                                 & 0.1670                           & 338.20                         & 25.12                           & 0.6343                           & 6.13                                 & 0.2384                           & 247.42                         \\
DRAN                               &                           & 32.80                           & 0.8756                           & 2.75                                 & 0.4337                           & 40.35                          & 35.36                           & 0.9574                           & 2.28                                 & 0.5527                           & 29.94                          & 34.08                           & 0.9165                           & 2.52                                 & 0.4932                           & 35.15                          \\
\textbf{Proposed} &                           & \textbf{41.09} & \textbf{0.9811} & \textbf{1.10}       & \textbf{0.671}  & \textbf{6.57} & \textbf{44.55} & \textbf{0.9927} & \textbf{0.69}       & \textbf{0.8175} & \textbf{3.21} & \textbf{42.82} & \textbf{0.9869} & \textbf{0.89}       & \textbf{0.7443} & \textbf{4.89} \\ \hline
CAE                                & \multirow{7}{*}{75}       & 32.85                           & 0.9105                           & 2.79                                 & 0.3433                           & 45.37                          & 33.97                           & 0.9382                           & 2.38                                 & 0.4065                           & 37.55                          & 33.41                           & 0.9244                           & 2.58                                 & 0.3749                           & 41.46                          \\
ResCNN                             &                           & 20.53                           & 0.2831                           & 12.52                                & 0.2101                           & 580.27                         & 25.53                           & 0.7212                           & 5.61                                 & 0.1956                           & 225.80                          & 23.03                           & 0.5022                           & 9.07                                 & 0.2029                           & 403.03                         \\
DnCNN                              &                           & 21.17                           & 0.3179                           & 11.06                                & 0.2214                           & 498.24                         & 27.00                           & 0.7713                           & 4.77                                 & 0.2361                           & 162.06                         & 24.09                           & 0.5446                           & 7.92                                 & 0.2288                           & 330.15                         \\
MMD                &                            & 28.45                           & 0.8304                           & 3.98                                 & 0.5082                           & 179.87                         & 28.45                           & 0.8304                           & 3.98                                 & 0.5082                           & 179.87                         & 28.45                           & 0.8304                           & 3.98                                 & 0.5082                           & 179.87                         \\
DAE                &                             & 21.13                           & 0.3085                           & 11.43                                & 0.2176                           & 505.25                         & 24.38                           & 0.6718                           & 6.33                                 & 0.1814                           & 300.59                         & 22.76                           & 0.4901                           & 8.88                                 & 0.1995                           & 402.92                         \\
DRAN                               &                           & 31.03                           & 0.7946                           & 3.03                                 & 0.3878                           & 57.14                          & 35.07                           & 0.9542                           & 2.35                                 & 0.5339                           & 31.53                          & 33.05                           & 0.8744                           & 2.69                                 & 0.4609                           & 44.34                          \\
\textbf{Proposed} &                           & \textbf{40.59} & \textbf{0.9787} & \textbf{1.15}       & \textbf{0.6581} & \textbf{7.30} & \textbf{43.07} & \textbf{0.9901} & \textbf{0.82}       & \textbf{0.7882} & \textbf{4.36} & \textbf{41.83} & \textbf{0.9844} & \textbf{0.99}       & \textbf{0.7232} & \textbf{5.83} 
\\ \hline
\end{tabular}}
\caption{Quantative comparison for MRI denoising. The proposed method outperforms existing denoising methods in Gaussian and speckle denoising by a notable margin. }
\label{MRITab}

\end{table*}

\subsubsection{X-Ray Image}
X-rays are widely used for diagnosing bone fractures, joint problems, lung conditions, dental issues, etc. Such radiography images are typically contaminated by additive (Gaussian) noise. Table \ref{xrayTab} illustrates the quantitative comparison between the deep denoising methods. The proposed method outperforms the existing denoising methods with 4.34 in PSNR,  0.0442 in SSIM, 0.97 in $\Delta E$, 0.0658 in VIFP, and  24.97 in MSE metric for speckle denoising on X-ray images. It depicts a similar improvement for Gaussian denoising by achieving a performance gain of 8.45 in PSNR, 0.1480 in SSIM, 2.77 in $\Delta E$, 0.2045 in VIFP, and 120.14 in MSE score.
 
\begin{figure*}[!htb]
\centering
\includegraphics[width=0.65\textwidth,keepaspectratio]{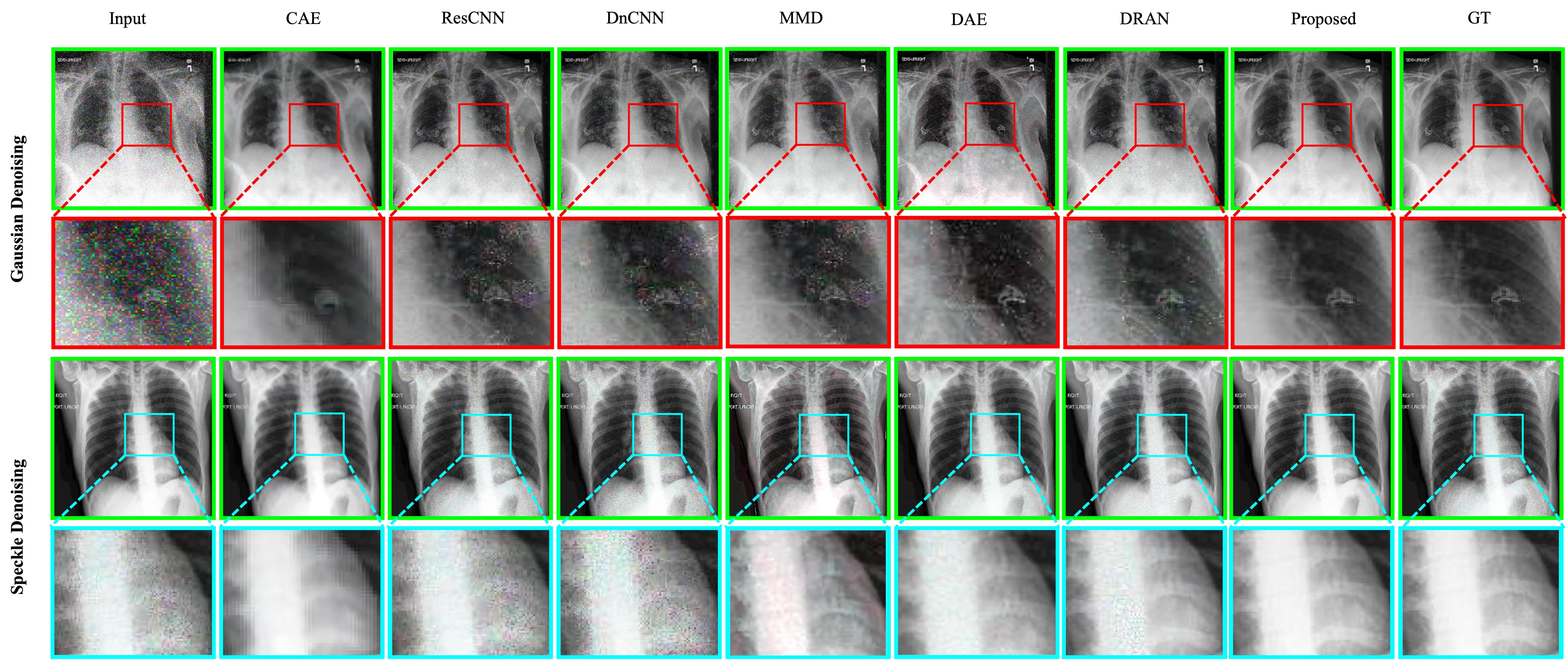}
\caption{ Qualitative comparison for X-ray denoising. The proposed method illustrates consistency in X-ray images as well. The top example shows Gaussian denoising; the bottom depicts speckle denoising. In each row, Left to right: Noisy input, CAE \cite{gondara2016medical}, ResCNN \cite{jifara2019medical}, DnCNN \cite{jiang2018denoising}, MMD \cite{el2022deep}, DAE \cite{el2022efficient}, DRAN \cite{sharif2020learning} \cite{sharif2020learning}, proposed method, and ground truth image.  }
\label{xray}
\end{figure*}

Fig. \ref{xray} visually confirms the performance gain achieved by the proposed two-stage network. Despite the denoising challenges, the proposed method illustrates the significant perceptual improvement in X-ray images for speckle and Gaussian denoising.

\begin{table*}[!htb]
\centering
\scalebox{.52}{
\begin{tabular}{lllllllllllllllll}
\hline
\multirow{2}{*}{\textbf{Model}}              & \multirow{2}{*}{\textbf{$\sigma$}}  & \multicolumn{5}{l}{\textbf{Gaussian}}                                                                                                                                                    & \multicolumn{5}{l}{\textbf{Speckle}}                                                                                                                                                     & \multicolumn{5}{l}{\textbf{Average}}                                                                                                                                                     \\ \cmidrule(){3-17}
                                   &                           & \textbf{PSNR}  & \textbf{SSIM}   & \textbf{$\Delta E$} & \textbf{VIFP}   & \textbf{MSE}   & \textbf{PSNR}  & \textbf{SSIM}   & \textbf{$\Delta E$} & \textbf{VIFP}   & \textbf{MSE}   & \textbf{PSNR}  & \textbf{SSIM}   & \textbf{$\Delta E$} & \textbf{VIFP}   & \textbf{MSE}   \\ \hline
CAE                                & \multirow{7}{*}{10}       & 30.43                           & 0.9178                           & 2.16                                 & 0.5049                           & 61.45                           & 30.25                           & 0.9181                           & 2.34                                 & 0.5061                           & 63.86                           & 30.34                           & 0.9179                           & 2.25                                 & 0.5055                           & 62.65                           \\
ResCNN                             &                           & 24.77                           & 0.7455                           & 4.90                                 & 0.3551                           & 241.32                          & 24.50                           & 0.7593                           & 4.63                                 & 0.4164                           & 247.90                          & 24.64                           & 0.7524                           & 4.77                                 & 0.3857                           & 244.61                          \\
DnCNN                              &                           & 26.19                           & 0.7812                           & 4.39                                 & 0.3879                           & 169.98                          & 24.29                           & 0.7317                           & 4.91                                 & 0.3925                           & 259.41                          & 25.24                           & 0.7564                           & 4.65                                 & 0.3902                           & 214.69                          \\
MMD                &                            & 23.43                           & 0.8288                           & 7.30                                 & 0.4307                           & 331.62                          & 23.43                           & 0.8288                           & 7.30                                 & 0.4307                           & 331.62                          & 23.43                           & 0.8288                           & 7.30                                 & 0.4307                           & 331.62                          \\
DAE                &                             & 27.63                           & 0.8537                           & 3.47                                 & 0.4620                           & 120.63                          & 26.67                           & 0.8264                           & 3.75                                 & 0.4796                           & 148.18                          & 27.15                           & 0.8401                           & 3.61                                 & 0.4708                           & 134.40                          \\
DRAN                               &                           & 33.35                           & 0.9236                           & 2.32                                 & 0.5919                           & 34.02                           & 35.11                           & 0.9500                           & 1.47                                 & 0.6443                           & 22.60                           & 34.23                           & 0.9368                           & 1.90                                 & 0.6181                           & 28.31                           \\
\textbf{Proposed} &                           & \textbf{37.32} & \textbf{0.9687} & \textbf{0.77}       & \textbf{0.6798} & \textbf{12.49} & \textbf{37.44} & \textbf{0.9695} & \textbf{0.76}       & \textbf{0.6840} & \textbf{12.16} & \textbf{37.38} & \textbf{0.9691} & \textbf{0.77}       & \textbf{0.6819} & \textbf{12.32} \\ \hline
CAE                                & \multirow{7}{*}{25}       & 30.51                           & 0.9150                           & 1.90                                 & 0.4965                           & 60.49                           & 30.50                           & 0.9181                           & 1.97                                 & 0.5050                           & 60.61                           & 30.50                           & 0.9166                           & 1.94                                 & 0.5007                           & 60.55                           \\
ResCNN                             &                           & 29.04                           & 0.8603                           & 2.94                                 & 0.4578                           & 87.95                           & 26.10                           & 0.8042                           & 3.79                                 & 0.4458                           & 176.23                          & 27.57                           & 0.8323                           & 3.36                                 & 0.4518                           & 132.09                          \\
DnCNN                              &                           & 28.01                           & 0.8299                           & 3.41                                 & 0.4190                           & 105.92                          & 25.44                           & 0.7741                           & 4.03                                 & 0.4285                           & 197.18                          & 26.73                           & 0.8020                           & 3.72                                 & 0.4237                           & 151.55                          \\
MMD                &                            & 23.46                           & 0.8307                           & 7.06                                 & 0.4239                           & 327.66                          & 23.46                           & 0.8307                           & 7.06                                 & 0.4239                           & 327.66                          & 23.46                           & 0.8307                           & 7.06                                 & 0.4239                           & 327.66                          \\
DAE                &                             & 30.28                           & 0.8749                           & 2.71                                 & 0.5160                           & 82.47                           & 29.58                           & 0.8777                           & 2.72                                 & 0.5422                           & 92.49                           & 29.93                           & 0.8763                           & 2.72                                 & 0.5291                           & 87.48                           \\
DRAN                               &                           & 30.84                           & 0.8828                           & 2.70                                 & 0.5336                           & 60.48                           & 33.18                           & 0.9267                           & 1.79                                 & 0.6167                           & 34.88                           & 32.01                           & 0.9047                           & 2.25                                 & 0.5751                           & 47.68                           \\
\textbf{Proposed} &                           & \textbf{37.13} & \textbf{0.9677} & \textbf{0.78}       & \textbf{0.6740} & \textbf{13.10} & \textbf{37.32} & \textbf{0.9687} & \textbf{0.77}       & \textbf{0.6795} & \textbf{12.50} & \textbf{37.22} & \textbf{0.9682} & \textbf{0.78}       & \textbf{0.6768} & \textbf{12.80} \\ \hline
CAE                                & \multirow{7}{*}{50}       & 30.22                           & 0.9071                           & 2.07                                 & 0.4763                           & 64.38                           & 30.53                           & 0.9159                           & 1.87                                 & 0.4983                           & 60.20                           & 30.38                           & 0.9115                           & 1.97                                 & 0.4873                           & 62.29                           \\
ResCNN                             &                           & 29.27                           & 0.8781                           & 2.52                                 & 0.4892                           & 77.30                           & 29.19                           & 0.8808                           & 2.64                                 & 0.5168                           & 86.12                           & 29.23                           & 0.8794                           & 2.58                                 & 0.5030                           & 81.71                           \\
DnCNN                              &                           & 28.10                           & 0.8335                           & 3.16                                 & 0.4175                           & 101.10                          & 26.77                           & 0.8137                           & 3.45                                 & 0.4558                           & 148.31                          & 27.44                           & 0.8236                           & 3.31                                 & 0.4366                           & 124.70                          \\
MMD                &                            & 23.84                           & 0.8323                           & 6.42                                 & 0.4189                           & 303.61                          & 23.84                           & 0.8323                           & 6.42                                 & 0.4189                           & 303.61                          & 23.84                           & 0.8323                           & 6.42                                 & 0.4189                           & 303.61                          \\
DAE                &                             & 27.94                           & 0.8589                           & 3.22                                 & 0.4628                           & 108.87                          & 32.67                           & 0.9199                           & 1.89                                 & 0.5910                           & 54.01                           & 30.30                           & 0.8894                           & 2.56                                 & 0.5269                           & 81.44                           \\
DRAN                               &                           & 26.27                           & 0.7800                           & 3.86                                 & 0.4229                           & 169.33                          & 32.53                           & 0.9229                           & 1.75                                 & 0.6103                           & 38.33                           & 29.40                           & 0.8515                           & 2.80                                 & 0.5166                           & 103.83                          \\
\textbf{Proposed} &                           & \textbf{36.95} & \textbf{0.9671} & \textbf{0.79}       & \textbf{0.6708} & \textbf{13.68} & \textbf{37.19} & \textbf{0.9681} & \textbf{0.78}       & \textbf{0.6758} & \textbf{12.91} & \textbf{37.07} & \textbf{0.9676} & \textbf{0.79}       & \textbf{0.6733} & \textbf{13.29} \\ \hline
CAE                                & \multirow{7}{*}{75}       & 29.95                           & 0.9000                           & 2.18                                 & 0.4598                           & 68.40                           & 30.47                           & 0.9136                           & 1.92                                 & 0.4917                           & 61.02                           & 30.21                           & 0.9068                           & 2.05                                 & 0.4757                           & 64.71                           \\
ResCNN                             &                           & 27.05                           & 0.8392                           & 3.50                                 & 0.4299                           & 129.19                          & 30.37                           & 0.9001                           & 2.41                                 & 0.5326                           & 63.78                           & 28.71                           & 0.8697                           & 2.95                                 & 0.4812                           & 96.49                           \\
DnCNN                              &                           & 26.53                           & 0.8146                           & 3.76                                 & 0.4050                           & 144.71                          & 27.71                           & 0.8367                           & 3.16                                 & 0.4732                           & 119.59                          & 27.12                           & 0.8256                           & 3.46                                 & 0.4391                           & 132.15                          \\
MMD                &                            & 23.82                           & 0.8275                           & 6.14                                 & 0.4116                           & 305.46                          & 23.82                           & 0.8275                           & 6.14                                 & 0.4116                           & 305.46                          & 23.82                           & 0.8275                           & 6.14                                 & 0.4116                           & 305.46                          \\
DAE                &                             & 26.09                           & 0.8039                           & 3.82                                 & 0.3846                           & 160.53                          & 33.01                           & 0.9223                           & 1.97                                 & 0.5861                           & 50.58                           & 29.55                           & 0.8631                           & 2.89                                 & 0.4853                           & 105.56                          \\
DRAN                               &                           & 23.89                           & 0.6918                           & 5.35                                 & 0.3270                           & 270.18                          & 30.86                           & 0.8978                           & 1.97                                 & 0.5786                           & 54.85                           & 27.38                           & 0.7948                           & 3.66                                 & 0.4528                           & 162.52                          \\
\textbf{Proposed} &                           & \textbf{36.79} & \textbf{0.9667} & \textbf{0.80}       & \textbf{0.6688} & \textbf{14.19} & \textbf{37.09} & \textbf{0.9677} & \textbf{0.78}       & \textbf{0.6737} & \textbf{13.22} & \textbf{36.94} & \textbf{0.9672} & \textbf{0.79}       & \textbf{0.6713} & \textbf{13.70} \\ \hline
\end{tabular}}
\caption{Quantative comparison for X-ray denoising. The proposed method outperforms existing denoising methods by a notable margin.}
\label{xrayTab}
\end{table*}

\subsubsection{CT Images}
CT scan is a type of medical imaging that uses X-rays and computer processing to create detailed information \cite{wu2020self}. To further confirm the performance of the proposed method in radiology images. Table. \ref{ctTab}, illustrates the quantitative comparison on CT images. The proposed method shows consistency in CT denoising and achieved a performance gain of over 4.34 in PSNR, 0.0442 in SSIM, 0.97 in $\Delta E$, 0.0658 in VIFP, and 24.97 in MSE metric for speckle denoising. Similarly, it outperforms the existing denoising methods by 8.46 in PSNR, 0.1480 in SSIM, 2.77 in $\Delta E$, 0.2045 in VIFP, and 120.14 in MSE metric for Gaussian denoising.

\begin{figure*}[!htb]
\centering
\includegraphics[width=0.65\textwidth,keepaspectratio]{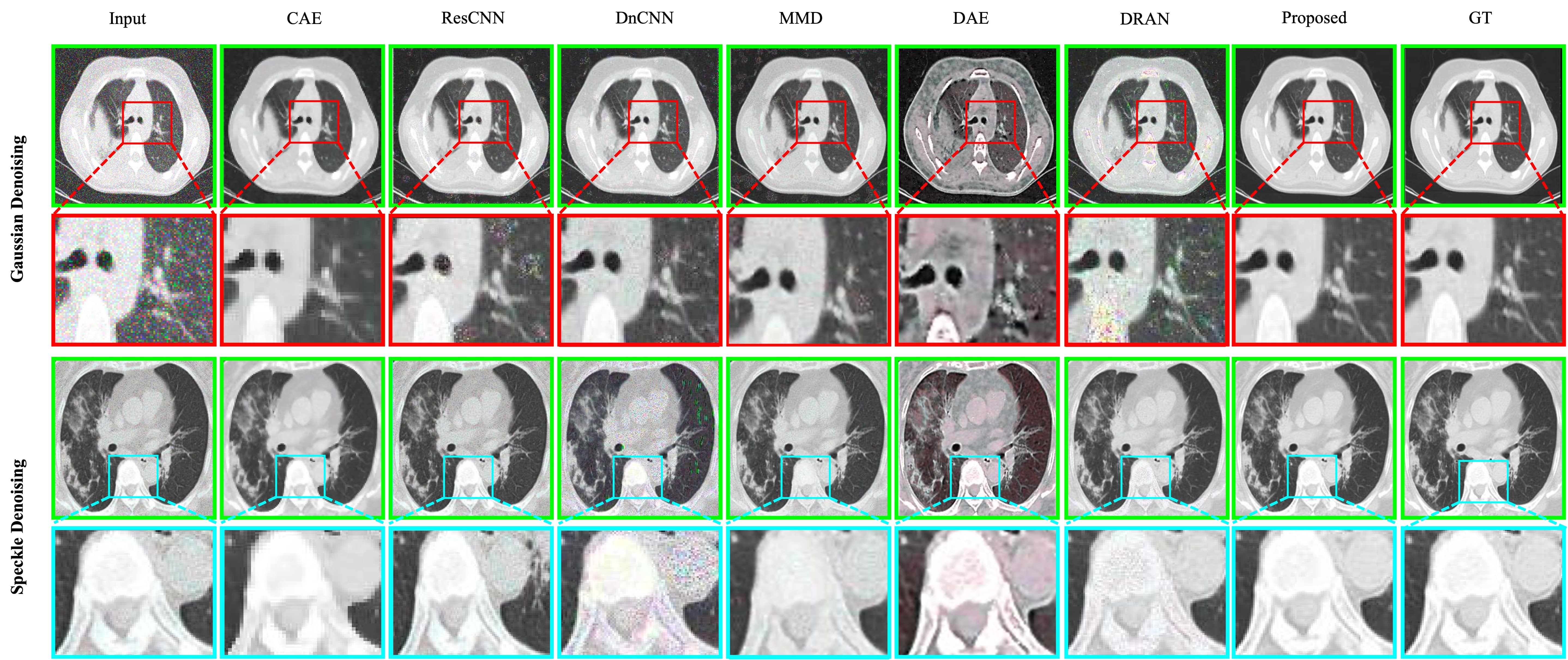}
\caption{ Qualitative comparison for CT denoising. The proposed method produces visually plausible images compared to the existing denoising methods. The top example depicts Gaussian denoising; the bottom illustrates speckle denoising. In each row, Left to right: Noisy input, CAE \cite{gondara2016medical}, ResCNN \cite{jifara2019medical}, DnCNN \cite{jiang2018denoising}, MMD \cite{el2022deep}, DAE \cite{el2022efficient}, DRAN \cite{sharif2020learning} \cite{sharif2020learning}, proposed method, and ground truth image. }
\label{ct}
\end{figure*}

Fig. \ref{ct} illustrates the visual comparison between the denoising methods. The proposed method can plausibly denoise the CT images as well. It can reduce noises from the CT images without illustrating any visual artifacts.

\begin{table*}[!htb]
\centering
\scalebox{.52}{\
\begin{tabular}{lclllllllllllllll}
\hline
\multirow{2}{*}{\textbf{Model}}              & \multirow{2}{*}{\textbf{$\sigma$}}      & \multicolumn{5}{l}{\textbf{Gaussian}}                                                                                                                                                   & \multicolumn{5}{l}{\textbf{Speckle}}                                                                                                                                                    & \multicolumn{5}{l}{\textbf{Average}}                                                                                                                                                    \\  \cmidrule(){3-17}
                                   &                            & \textbf{PSNR}  & \textbf{SSIM}   & \textbf{$\Delta E$} & \textbf{VIFP}   & \textbf{MSE}  & \textbf{PSNR}  & \textbf{SSIM}   & \textbf{$\Delta E$} & \textbf{VIFP}   & \textbf{MSE}  & \textbf{PSNR}  & \textbf{SSIM}   & \textbf{$\Delta E$} & \textbf{VIFP}   & \textbf{MSE}  \\ \hline
CAE                                & \multirow{7}{*}{10}                          & 27.39                           & 0.8882                           & 2.39                                 & 0.4311                           & 138.61                         & 27.30                           & 0.8886                           & 2.49                                 & 0.4316                           & 141.21                         & 27.34                           & 0.8884                           & 2.44                                 & 0.4314                           & 139.91                         \\
ResCNN                             &                            & 23.92                           & 0.7214                           & 7.43                                 & 0.3771                           & 286.38                         & 22.89                           & 0.7051                           & 7.67                                 & 0.3745                           & 358.90                         & 23.41                           & 0.7133                           & 7.55                                 & 0.3758                           & 322.64                         \\
DnCNN                              &                            & 23.29                           & 0.6763                           & 8.02                                 & 0.3465                           & 347.46                         & 22.28                           & 0.6604                           & 8.09                                 & 0.3452                           & 427.39                         & 22.79                           & 0.6684                           & 8.05                                 & 0.3459                           & 387.43                         \\
MMD                &                            & 18.60                           & 0.8004                           & 9.89                                 & 0.4403                           & 1135.39                        & 18.60                           & 0.8004                           & 9.89                                 & 0.4403                           & 1135.39                        & 18.60                           & 0.8004                           & 9.89                                 & 0.4403                           & 1135.39                        \\
DAE                &                              & 25.05                           & 0.7498                           & 6.89                                 & 0.4074                           & 238.80                         & 24.38                           & 0.7460                           & 6.95                                 & 0.4130                           & 268.49                         & 24.72                           & 0.7479                           & 6.92                                 & 0.4102                           & 253.65                         \\
DRAN                               &                            & 36.72                           & 0.9624                           & 1.64                                 & 0.7226                           & 17.48                          & 37.54                           & 0.9699                           & 1.39                                 & 0.7554                           & 14.31                          & 37.13                           & 0.9662                           & 1.52                                 & 0.7390                           & 15.90                          \\
\textbf{Proposed} &                            & \textbf{42.20} & \textbf{0.9869} & \textbf{0.59}       & \textbf{0.8303} & \textbf{5.64} & \textbf{42.93} & \textbf{0.9880}  & \textbf{0.54}       & \textbf{0.8471} & \textbf{5.15} & \textbf{42.56} & \textbf{0.9875} & \textbf{0.57}       & \textbf{0.8387} & \textbf{5.40} \\ \hline
CAE                                & \multirow{7}{*}{25}                          & 27.42                           & 0.8853                           & 2.26                                 & 0.4259                           & 138.14                         & 27.41                           & 0.8883                           & 2.28                                 & 0.4311                           & 138.42                         & 27.42                           & 0.8868                           & 2.27                                 & 0.4285                           & 138.28                         \\
ResCNN                             &                            & 26.50                           & 0.8010                           & 5.71                                 & 0.4435                           & 167.40                         & 24.34                           & 0.7540                           & 6.69                                 & 0.4076                           & 270.43                         & 25.42                           & 0.7775                           & 6.20                                 & 0.4256                           & 218.92                         \\
DnCNN                              &                            & 25.22                           & 0.7448                           & 6.84                                 & 0.3907                           & 224.91                         & 23.36                           & 0.7092                           & 7.35                                 & 0.3767                           & 333.93                         & 24.29                           & 0.7270                           & 7.10                                 & 0.3837                           & 279.42                         \\
MMD                &                            & 18.80                           & 0.8014                           & 9.63                                 & 0.4337                           & 1099.53                        & 18.80                           & 0.8014                           & 9.63                                 & 0.4337                           & 1099.53                        & 18.80                           & 0.8014                           & 9.63                                 & 0.4337                           & 1099.53                        \\
DAE                &                              & 26.84                           & 0.7929                           & 5.70                                 & 0.4494                           & 174.33                         & 25.20                           & 0.7718                           & 6.37                                 & 0.4324                           & 243.72                         & 26.02                           & 0.7823                           & 6.04                                 & 0.4409                           & 209.03                         \\
DRAN                               &                            & 32.01                           & 0.8950                           & 3.13                                 & 0.5758                           & 62.51                          & 35.50                           & 0.9500                             & 1.85                                 & 0.6961                           & 24.13                          & 33.76                           & 0.9225                           & 2.49                                 & 0.6360                           & 43.32                          \\
\textbf{Proposed} & \textbf{} & \textbf{40.99} & \textbf{0.9847} & \textbf{0.64}       & \textbf{0.8027} & \textbf{6.96} & \textbf{42.02} & \textbf{0.9864} & \textbf{0.58}       & \textbf{0.8253} & \textbf{5.83} & \textbf{41.51} & \textbf{0.9856} & \textbf{0.61}       & \textbf{0.8140} & \textbf{6.40} \\ \hline
CAE                                & \multirow{7}{*}{50}                          & 27.27                           & 0.8778                           & 2.35                                 & 0.4136                           & 142.23                         & 27.43                           & 0.8862                           & 2.24                                 & 0.4264                           & 138.15                         & 27.35                           & 0.8820                           & 2.29                                 & 0.4200                           & 140.19                         \\
ResCNN                             &                            & 27.65                           & 0.8506                           & 4.23                                 & 0.4986                           & 116.56                         & 25.83                           & 0.7992                           & 5.84                                 & 0.4427                           & 195.83                         & 26.74                           & 0.8249                           & 5.03                                 & 0.4707                           & 156.20                         \\
DnCNN                              &                            & 26.55                           & 0.8134                           & 5.29                                 & 0.4509                           & 155.26                         & 25.07                           & 0.7691                           & 6.39                                 & 0.4181                           & 227.94                         & 25.81                           & 0.7913                           & 5.84                                 & 0.4345                           & 191.60                         \\
MMD                &                            & 18.97                           & 0.7974                           & 9.40                                 & 0.4209                           & 1069.68                        & 18.97                           & 0.7974                           & 9.40                                 & 0.4209                           & 1069.68                        & 18.97                           & 0.7974                           & 9.40                                 & 0.4209                           & 1069.68                        \\
DAE                &                              & 25.67                           & 0.7745                           & 5.91                                 & 0.4354                           & 250.16                         & 26.03                           & 0.7916                           & 6.07                                 & 0.4491                           & 228.54                         & 25.85                           & 0.7830                           & 5.99                                 & 0.4423                           & 239.35                         \\
DRAN                               &                            & 27.94                           & 0.8229                           & 4.84                                 & 0.4468                           & 130.15                         & 33.23                           & 0.9203                           & 2.51                                 & 0.6217                           & 40.11                          & 30.59                           & 0.8716                           & 3.68                                 & 0.5343                           & 85.13                          \\
\textbf{Proposed} &                            & \textbf{40.06} & \textbf{0.983}  & \textbf{0.69}       & \textbf{0.7848} & \textbf{8.47} & \textbf{41.12} & \textbf{0.9850}  & \textbf{0.62}       & \textbf{0.8057} & \textbf{6.79} & \textbf{40.59} & \textbf{0.9840} & \textbf{0.66}       & \textbf{0.7953} & \textbf{7.63} \\ \hline
CAE                                & \multirow{7}{*}{75}                        & 27.09                           & 0.8709                           & 2.43                                 & 0.4036                           & 147.23                         & 27.40                           & 0.8842                           & 2.26                                 & 0.4222                           & 138.77                         & 27.25                           & 0.8776                           & 2.34                                 & 0.4129                           & 143.00                         \\
ResCNN                             &                            & 26.57                           & 0.8261                           & 4.98                                 & 0.4677                           & 158.42                         & 27.30                           & 0.8390                           & 5.11                                 & 0.4792                           & 144.05                         & 26.94                           & 0.8326                           & 5.05                                 & 0.4735                           & 151.24                         \\
DnCNN                              &                            & 25.61                           & 0.8176                           & 5.31                                 & 0.4586                           & 188.14                         & 26.54                           & 0.8112                           & 5.61                                 & 0.4538                           & 166.93                         & 26.08                           & 0.8144                           & 5.46                                 & 0.4562                           & 177.53                         \\
MMD                &                            & 18.97                           & 0.7894                           & 9.37                                 & 0.4084                           & 1071.27                        & 18.97                           & 0.7894                           & 9.37                                 & 0.4084                           & 1071.27                        & 18.97                           & 0.7894                           & 9.37                                 & 0.4084                           & 1071.27                        \\
DAE                &                              & 25.21                           & 0.7915                           & 5.60                                 & 0.4289                           & 260.84                         & 26.76                           & 0.8047                           & 5.81                                 & 0.4645                           & 219.53                         & 25.98                           & 0.7981                           & 5.71                                 & 0.4467                           & 240.18                         \\
DRAN                               &                            & 27.32                           & 0.8158                           & 4.98                                 & 0.4313                           & 141.79                         & 31.66                           & 0.8949                           & 3.08                                 & 0.5686                           & 59.29                          & 29.49                           & 0.8554                           & 4.03                                 & 0.5000                           & 100.54                         \\
\textbf{Proposed} &                            & \textbf{39.50} & \textbf{0.9819} & \textbf{0.74}       & \textbf{0.7752} & \textbf{9.58} & \textbf{40.55} & \textbf{0.9842} & \textbf{0.64}       & \textbf{0.7949} & \textbf{7.60} & \textbf{40.03} & \textbf{0.9831} & \textbf{0.69}       & \textbf{0.7851} & \textbf{8.59} \\ \hline
\end{tabular}}
\caption{Quantative comparison for CT denoising. The proposed method demonstrates consistency in CT denoising as well.}
\label{ctTab}
\end{table*}

\subsubsection{Skin Images}
Skin lesion images are medical images that show various skin abnormalities, including moles, freckles, rashes, and tumors. Such medical image types are contaminated with numerous noise patterns. This study extensively evaluated our proposed two-stage network for multi-pattern denoising. Table. \ref{hrmTab} illustrates the quantitative comparison for Gaussian and speckle denoising. The proposed method achieved significant improvement in skin image denoising. The proposed method depicts an improvement for speckle denoising by improving  8.51 in PSNR,  0.1184 in SSIM, 1.34 in $\Delta E$, 0.2214 in VIFP, and  28.20 in MSE score. The proposed method also illustrates a performance gain of  10.68 in PSNR, 0.1868 in SSIM, 2.20 in $\Delta E$, 0.3005 in VIFP, and 59.33 in MSE for Gaussian denoising. 

\begin{figure*}[!htb]
\centering
\includegraphics[width=0.65\textwidth,keepaspectratio]{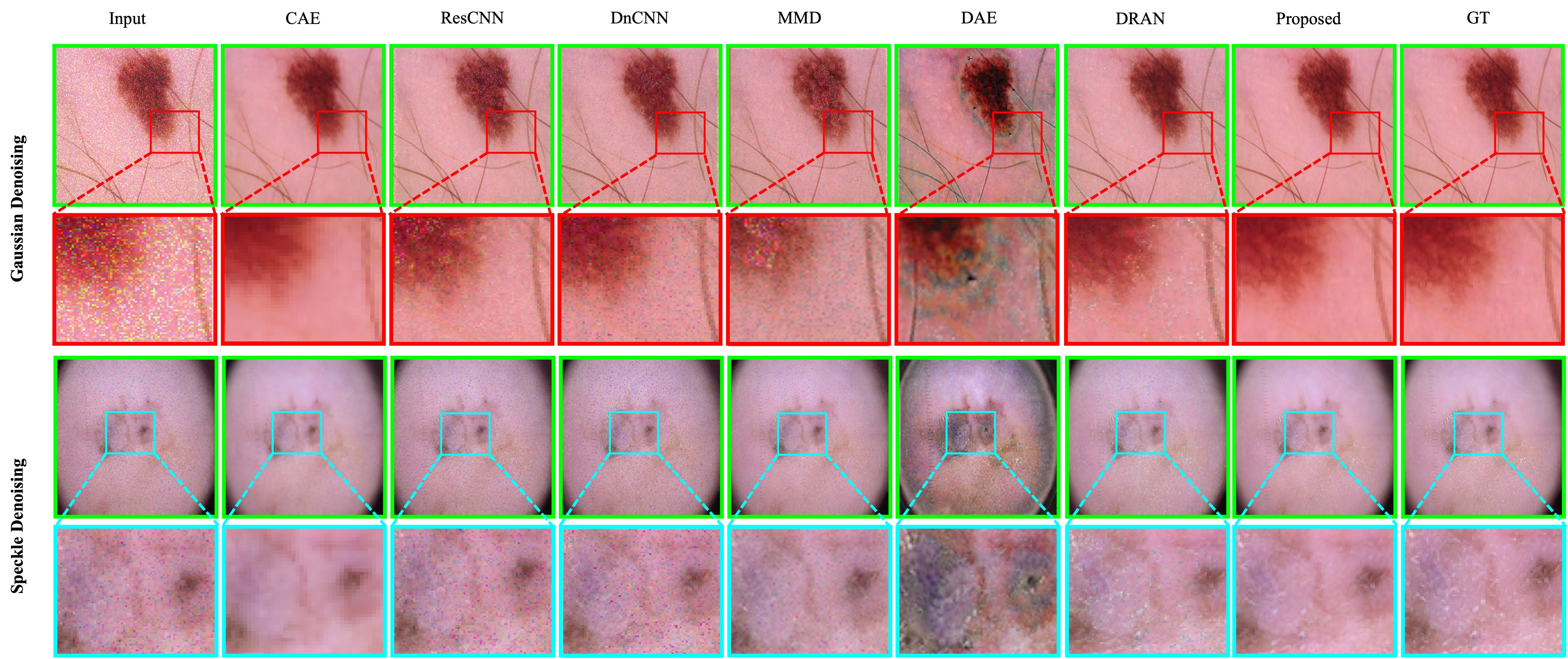}
\caption{ Qualitative comparison for skin image denoising. The proposed method outperforms the existing denoising methods by generating plausible images. The top example illustrates Gaussian denoising; the bottom demonstrates speckle denoising. In each row, Left to right: Noisy input, CAE \cite{gondara2016medical}, ResCNN \cite{jifara2019medical}, DnCNN \cite{jiang2018denoising}, MMD \cite{el2022deep}, DAE \cite{el2022efficient}, DRAN \cite{sharif2020learning} \cite{sharif2020learning}, proposed method, and ground truth image. }
\label{hrm}
\end{figure*}

In addition to the quantitative comparison, we extensively evaluated the deep models with visual results, as shown in Fig. \ref{hrm}. It is visible that the proposed method can produce clean medical images without any visually disturbing artifacts for speckle and Gaussian denoising.

\begin{table*}[!htb]
\centering
\scalebox{.52}{
\begin{tabular}{lllllllllllllllll}
\hline
\multirow{2}{*}{\textbf{Model}}           & \multirow{2}{*}{\textbf{$\sigma$}}  & \multicolumn{5}{l}{\textbf{Gaussian}}                                                                                                                                                   & \multicolumn{5}{l}{\textbf{Speckle}}                                                                                                                                                    & \multicolumn{5}{l}{\textbf{Average}}                                                                                                                                                    \\ \cmidrule(){3-17}
\cmidrule(){3-17} &                           & \textbf{PSNR}  & \textbf{SSIM}   & \textbf{$\Delta E$} & \textbf{VIFP}   & \textbf{MSE}  & \textbf{PSNR}  & \textbf{SSIM}   & \textbf{$\Delta E$} & \textbf{VIFP}   & \textbf{MSE}  & \textbf{PSNR}  & \textbf{SSIM}   & \textbf{$\Delta E$} & \textbf{VIFP}   & \textbf{MSE}  \\ \hline
CAE                                & \multirow{7}{*}{10}       & 34.84                           & 0.9209                           & 1.84                                 & 0.4493                           & 25.22                          & 34.57                           & 0.9220                           & 1.88                                 & 0.4533                           & 26.64                          & 34.70                           & 0.9214                           & 1.86                                 & 0.4513                           & 25.93                          \\
ResCNN                             &                           & 21.41                           & 0.3084                           & 9.49                                 & 0.1549                           & 503.15                         & 20.59                           & 0.2827                           & 10.57                                & 0.1499                           & 599.68                         & 21.00                           & 0.2955                           & 10.03                                & 0.1524                           & 551.41                         \\
DnCNN                              &                           & 20.53                           & 0.2655                           & 10.61                                & 0.1464                           & 603.55                         & 20.17                           & 0.2596                           & 11.01                                & 0.1513                           & 653.57                         & 20.35                           & 0.2626                           & 10.81                                & 0.1488                           & 628.56                         \\
MMD                &                            & 20.41                           & 0.7221                           & 8.61                                 & 0.4830                           & 678.19                         & 20.41                           & 0.7221                           & 8.61                                 & 0.4830                           & 678.19                         & 20.41                           & 0.7221                           & 8.61                                 & 0.4830                           & 678.19                         \\
DAE                &                             & 22.26                           & 0.3819                           & 8.98                                 & 0.1885                           & 447.88                         & 22.15                           & 0.3810                           & 9.11                                 & 0.1969                           & 439.06                         & 22.21                           & 0.3814                           & 9.04                                 & 0.1927                           & 443.47                         \\
DRAN                               &                           & 37.35                           & 0.9350                            & 1.54                                 & 0.6006                           & 13.41                          & 38.48                           & 0.9488                           & 1.35                                 & 0.6587                           & 10.75                          & 37.92                           & 0.9419                           & 1.45                                 & 0.6296                           & 12.08                          \\
\textbf{Proposed} &                           & \textbf{42.64} & \textbf{0.9788} & \textbf{0.89}       & \textbf{0.7393} & \textbf{3.97} & \textbf{43.43} & \textbf{0.9814} & \textbf{0.84}       & \textbf{0.7649} & \textbf{3.29} & \textbf{43.04} & \textbf{0.9801} & \textbf{0.87}       & \textbf{0.7521} & \textbf{3.63} \\ \hline
CAE                                & \multirow{7}{*}{25}       & 34.84                           & 0.9151                           & 1.84                                 & 0.4312                           & 24.94                          & 34.92                           & 0.9198                           & 1.81                                 & 0.4472                           & 24.77                          & 34.88                           & 0.9174                           & 1.83                                 & 0.4392                           & 24.85                          \\
ResCNN                             &                           & 24.04                           & 0.4300                           & 7.02                                 & 0.1941                           & 314.55                         & 22.40                           & 0.3600                             & 8.45                                 & 0.1840                            & 420.60                          & 23.22                           & 0.3950                            & 7.73                                 & 0.1890                            & 367.58                         \\
DnCNN                              &                           & 22.37                           & 0.3358                           & 8.51                                 & 0.1625                           & 385.6                          & 20.93                           & 0.2998                           & 9.83                                 & 0.1658                           & 542.33                         & 21.65                           & 0.3178                           & 9.17                                 & 0.1641                           & 463.96                         \\
MMD                &                            & 20.15                           & 0.7071                           & 8.88                                 & 0.4394                           & 732.22                         & 20.15                           & 0.7071                           & 8.88                                 & 0.4394                           & 732.22                         & 20.15                           & 0.7071                           & 8.88                                 & 0.4394                           & 732.22                         \\
DAE                &                             & 24.96                           & 0.5147                           & 6.63                                 & 0.2391                           & 341.52                         & 22.91                           & 0.4192                           & 8.31                                 & 0.2123                           & 419.90                         & 23.94                           & 0.4669                           & 7.47                                 & 0.2257                           & 380.71                         \\
DRAN                               &                           & 33.85                           & 0.8663                           & 2.20                                 & 0.4802                           & 28.46                          & 36.46                           & 0.9144                           & 1.67                                 & 0.5739                           & 16.24                          & 35.15                           & 0.8903                           & 1.94                                 & 0.5270                            & 22.35                          \\
\textbf{Proposed} &                           & \textbf{42.61} & \textbf{0.9776} & \textbf{0.89}       & \textbf{0.7306} & \textbf{4.11} & \textbf{43.1}  & \textbf{0.9790} & \textbf{0.86}       & \textbf{0.7416} & \textbf{3.58} & \textbf{42.85} & \textbf{0.9783} & \textbf{0.87}       & \textbf{0.7361} & \textbf{3.84} \\ \hline
CAE                                & \multirow{7}{*}{50}       & 34.19                           & 0.904                            & 1.99                                 & 0.3993                           & 28.07                          & 34.73                           & 0.9141                           & 1.84                                 & 0.4291                           & 25.34                          & 34.46                           & 0.9090                            & 1.91                                 & 0.4142                           & 26.71                          \\
ResCNN                             &                           & 25.60                           & 0.6005                           & 5.42                                 & 0.2613                           & 224.78                         & 25.00                           & 0.4871                           & 6.33                                 & 0.2283                           & 263.92                         & 25.30                            & 0.5438                           & 5.87                                 & 0.2448                           & 244.35                         \\
DnCNN                              &                           & 25.84                           & 0.5778                           & 5.20                                 & 0.2454                           & 176.86                         & 23.25                           & 0.3983                           & 7.47                                 & 0.1979                           & 316.48                         & 24.55                           & 0.4880                           & 6.34                                 & 0.2216                           & 246.67                         \\
MMD                &                            & 19.62                           & 0.6885                           & 9.54                                 & 0.3886                           & 837.40                         & 19.62                           & 0.6885                           & 9.54                                 & 0.3886                           & 837.40                         & 19.62                           & 0.6885                           & 9.54                                 & 0.3886                           & 837.40                         \\
DAE                &                             & 25.54                           & 0.5740                           & 5.04                                 & 0.2359                           & 194.36                         & 26.61                           & 0.5778                           & 5.55                                 & 0.2775                           & 259.65                         & 26.08                           & 0.5759                           & 5.30                                 & 0.2567                           & 227.01                         \\
DRAN                               &                           & 29.23                           & 0.7201                           & 3.73                                 & 0.3513                           & 80.50                          & 32.84                           & 0.8292                           & 2.46                                 & 0.4641                           & 36.32                          & 31.04                           & 0.7747                           & 3.10                                 & 0.4077                           & 58.41                          \\
\textbf{Proposed} &                           & \textbf{42.56} & \textbf{0.9775} & \textbf{0.89}       & \textbf{0.7315} & \textbf{4.23} & \textbf{42.9}  & \textbf{0.9783} & \textbf{0.87}       & \textbf{0.7358} & \textbf{3.82} & \textbf{42.73} & \textbf{0.9779} & \textbf{0.88}       & \textbf{0.7337} & \textbf{4.02} \\ \hline
CAE                                & \multirow{7}{*}{75}       & 33.87                           & 0.8941                           & 2.09                                 & 0.3748                           & 29.94                          & 34.46                           & 0.9084                           & 1.90                                 & 0.4123                           & 26.52                          & 34.17                           & 0.9012                           & 2.00                                 & 0.3936                           & 28.23                          \\
ResCNN                             &                           & 23.45                           & 0.5201                           & 7.70                                 & 0.2183                           & 312.98                         & 27.20                           & 0.6096                           & 4.92                                 & 0.2784                           & 179.16                         & 25.32                           & 0.5649                           & 6.31                                 & 0.2484                           & 246.07                         \\
DnCNN                              &                           & 24.41                           & 0.5516                           & 6.72                                 & 0.2175                           & 240.72                         & 25.55                           & 0.5200                             & 5.60                                 & 0.2401                           & 190.39                         & 24.98                           & 0.5358                           & 6.16                                 & 0.2288                           & 215.56                         \\
MMD                &                            & 19.15                           & 0.6732                           & 10.12                                & 0.3546                           & 929.46                         & 19.15                           & 0.6732                           & 10.12                                & 0.3546                           & 929.46                         & 19.15                           & 0.6732                           & 10.12                                & 0.3546                           & 929.46                         \\
DAE                &                             & 24.37                           & 0.5769                           & 6.06                                 & 0.2194                           & 240.96                         & 27.77                           & 0.6263                           & 4.31                                 & 0.2884                           & 130.90                         & 26.07                           & 0.6016                           & 5.19                                 & 0.2539                           & 185.93                         \\
DRAN                               &                           & 27.08                           & 0.6423                           & 4.89                                 & 0.2964                           & 131.62                         & 30.37                           & 0.7506                           & 3.33                                 & 0.3947                           & 64.16                          & 28.72                           & 0.6964                           & 4.11                                 & 0.3456                           & 97.89                          \\
\textbf{Proposed} &                           & \textbf{42.40}  & \textbf{0.9771} & \textbf{0.90}       & \textbf{0.7292} & \textbf{4.38} & \textbf{42.77} & \textbf{0.9780} & \textbf{0.88}       & \textbf{0.7346} & \textbf{3.98} & \textbf{42.59} & \textbf{0.9775} & \textbf{0.89}       & \textbf{0.7319} & \textbf{4.18} \\ \hline
\end{tabular}}
\caption{Quantitative comparison for skin image denoising. The proposed method can equally denoise the skin images.}
\label{hrmTab}
\end{table*}

\subsubsection{Microscopy Images}
Microscopy plays a crucial role in medical imaging as it allows researchers and medical professionals to study the structure and function of cells, tissues, and organs in the human body at a microscopic level. Despite having a substantial impact, these also can be affected by numerous noise patterns. Table. \ref{micTab} illustrates the quantitative comparison for the microscopic images. The proposed method can improve the cumulative performance of 5.79 in PSNR, 0.1184 in SSIM, 0.24 in $\Delta E$, 0.1045 in VIFP, and 11.825 in MSE metric for speckle denoising. It also illustrates a performance gain of  6.59 in PSNR,  0.0805 in SSIM, 0.87 in $\Delta E$,  0.1742 in VIFP, and 25.03 in MSE  metric for Gaussian denoising. 

\begin{figure*}[!htb]
\centering
\includegraphics[width=0.65\textwidth,keepaspectratio]{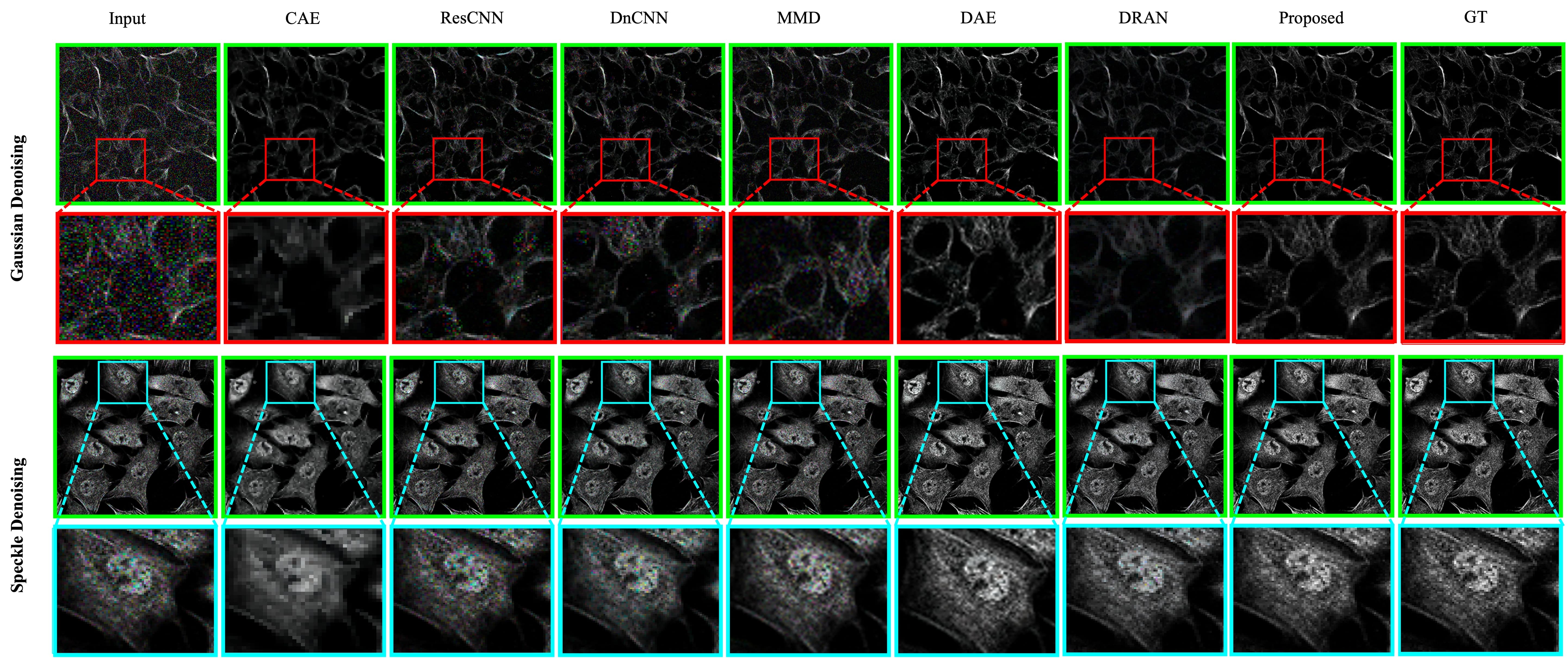}
\caption{Qualitative comparison for microscopy image denoising. The proposed method generates visually plausible images without illustrating visual artifacts. The top example illustrates Gaussian denoising;  the bottom illustrates speckle denoising. In each row, Left to right: Noisy input, CAE \cite{gondara2016medical}, ResCNN \cite{jifara2019medical}, DnCNN \cite{jiang2018denoising}, MMD \cite{el2022deep}, DAE \cite{el2022efficient}, DRAN \cite{sharif2020learning} \cite{sharif2020learning}, proposed method, and ground truth image. }
\label{micro}
\end{figure*}

Similar to radiology and skin images, the proposed method illustrates significant performance gain over the existing denoising methods for microscopic images. Fig. \ref{micro} further confirms the performance gain of the proposed method by qualitative results. Overall, the proposed method illustrates consistent all comparing modalities in qualitative and quantitative comparisons. 
 
		-		-

\begin{table*}[!htb]
\centering
\scalebox{.52}{
\begin{tabular}{lllllllllllllllll}
\hline
\multirow{2}{*}{\textbf{Model}}              & \multirow{2}{*}{\textbf{$\sigma$}}  & \multicolumn{5}{l}{\textbf{Gaussian}}                                                                                                                                                    & \multicolumn{5}{l}{\textbf{Speckle}}                                                                                                                                                    & \multicolumn{5}{l}{\textbf{Average}}                                                                                                                                                     \\ \cmidrule(){3-17}
                                   &                           & \textbf{PSNR}  & \textbf{SSIM}   & \textbf{$\Delta E$} & \textbf{VIFP}   & \textbf{MSE}   & \textbf{PSNR}  & \textbf{SSIM}   & \textbf{$\Delta E$} & \textbf{VIFP}   & \textbf{MSE}  & \textbf{PSNR}  & \textbf{SSIM}   & \textbf{$\Delta E$} & \textbf{VIFP}   & \textbf{MSE}   \\ \hline
CAE                                & \multirow{7}{*}{10}       & 32.07                           & 0.9094                           & 0.98                                 & 0.3726                           & 79.23                           & 32.01                           & 0.9087                           & 0.97                                 & 0.3730                           & 80.68                          & 32.04                           & 0.9091                           & 0.98                                 & 0.3728                           & 79.96                           \\
ResCNN                             &                           & 30.64                           & 0.8710                           & 2.44                                 & 0.3163                           & 83.31                           & 28.10                           & 0.8294                           & 2.68                                 & 0.2272                           & 175.00                         & 29.37                           & 0.8502                           & 2.56                                 & 0.2718                           & 129.15                          \\
DnCNN                              &                           & 30.34                           & 0.8660                           & 2.43                                 & 0.3216                           & 90.11                           & 28.11                           & 0.8132                           & 2.43                                 & 0.2432                           & 158.22                         & 29.23                           & 0.8396                           & 2.43                                 & 0.2824                           & 124.17                          \\
MMD                &                            & 26.78                           & 0.7327                           & 3.16                                 & 0.5058                           & 385.58                          & 26.78                           & 0.7327                           & 3.16                                 & 0.5058                           & 385.58                         & 26.78                           & 0.7327                           & 3.16                                 & 0.5058                           & 385.58                          \\
DAE                &                             & 29.55                           & 0.8514                           & 2.80                                 & 0.2996                           & 108.10                          & 27.22                           & 0.8262                           & 2.76                                 & 0.2225                           & 174.56                         & 28.39                           & 0.8388                           & 2.78                                 & 0.2610                           & 141.33                          \\
DRAN                               &                           & 37.11                           & 0.9693                           & 0.67                                 & 0.5904                           & 19.09                           & 38.63                           & 0.9859                           & 0.46                                 & 0.6900                             & 15.12                          & 37.87                           & 0.9776                           & 0.56                                 & 0.6402                           & 17.11                           \\
\textbf{Proposed} &                           & \textbf{43.14} & \textbf{0.9888} & \textbf{0.29}       & \textbf{0.7415} & \textbf{5.74}  & \textbf{44.35} & \textbf{0.9927} & \textbf{0.25}       & \textbf{0.7848} & \textbf{4.93} & \textbf{43.74} & \textbf{0.9908} & \textbf{0.27}       & \textbf{0.7632} & \textbf{5.34}  \\ \hline
CAE                                & \multirow{7}{*}{25}       & 31.95                           & 0.9053                           & 0.99                                 & 0.3640                            & 80.12                           & 32.05                           & 0.9096                           & 0.97                                 & 0.3749                           & 79.69                          & 32.00                           & 0.9075                           & 0.98                                 & 0.3695                           & 79.91                           \\
ResCNN                             &                           & 30.63                           & 0.8449                           & 2.57                                 & 0.3548                           & 65.50                           & 28.33                           & 0.8356                           & 2.65                                 & 0.2340                            & 163.00                         & 29.48                           & 0.8403                           & 2.61                                 & 0.2944                           & 114.25                          \\
DnCNN                              &                           & 29.95                           & 0.8541                           & 2.70                                 & 0.3361                           & 79.96                           & 28.38                           & 0.8236                           & 2.47                                 & 0.2573                           & 148.05                         & 29.16                           & 0.8389                           & 2.59                                 & 0.2967                           & 114.01                          \\
MMD                &                            & 26.62                           & 0.7290                           & 3.20                                 & 0.5015                           & 388.86                          & 26.62                           & 0.7290                           & 3.20                                 & 0.5015                           & 388.86                         & 26.62                           & 0.7290                           & 3.20                                 & 0.5015                           & 388.86                          \\
DAE                &                             & 31.14                           & 0.8759                           & 2.49                                 & 0.3877                           & 58.51                           & 27.39                           & 0.8345                           & 2.71                                 & 0.2315                           & 165.22                         & 29.27                           & 0.8552                           & 2.60                                 & 0.3096                           & 111.87                          \\
DRAN                               &                           & 34.91                           & 0.9408                           & 0.98                                 & 0.5186                           & 29.30                           & 38.36                           & 0.9852                           & 0.48                                 & 0.6823                           & 16.01                          & 36.63                           & 0.9630                           & 0.73                                 & 0.6005                           & 22.66                           \\
\textbf{Proposed} &                           & \textbf{41.33} & \textbf{0.9832} & \textbf{0.34}       & \textbf{0.6899} & \textbf{8.20}   & \textbf{44.19} & \textbf{0.9926} & \textbf{0.26}       & \textbf{0.7825} & \textbf{5.06} & \textbf{42.76} & \textbf{0.9879} & \textbf{0.30}       & \textbf{0.7362} & \textbf{6.63}  \\ \hline
CAE                                & \multirow{7}{*}{50}       & 31.73                           & 0.8993                           & 1.02                                 & 0.3514                           & 82.96                           & 32.08                           & 0.9103                           & 0.97                                 & 0.3763                           & 79.29                          & 31.9                            & 0.9048                           & 0.99                                 & 0.3639                           & 81.13                           \\
ResCNN                             &                           & 27.06                           & 0.6417                           & 4.36                                 & 0.3332                           & 129.69                          & 28.67                           & 0.8453                           & 2.59                                 & 0.2451                           & 144.51                         & 27.87                           & 0.7435                           & 3.48                                 & 0.2892                           & 137.10                          \\
DnCNN                              &                           & 27.2                            & 0.7543                           & 3.29                                 & 0.3370                            & 126.55                          & 28.65                           & 0.8384                           & 2.46                                 & 0.2708                           & 131.70                         & 27.93                           & 0.7964                           & 2.87                                 & 0.3039                           & 129.13                          \\
MMD                &                            & 26.23                           & 0.7200                           & 3.29                                 & 0.4849                           & 404.30                          & 26.23                           & 0.7200                           & 3.29                                 & 0.4849                           & 404.30                         & 26.23                           & 0.7200                           & 3.29                                 & 0.4849                           & 404.30                          \\
DAE                &                             & 26.80                           & 0.6347                           & 4.70                                 & 0.3271                           & 140.48                          & 27.42                           & 0.8384                           & 2.66                                 & 0.2371                           & 161.80                         & 27.11                           & 0.7365                           & 3.68                                 & 0.2821                           & 151.14                          \\
DRAN                               &                           & 33.00                           & 0.8785                           & 1.47                                 & 0.4692                           & 41.63                           & 37.85                           & 0.9836                           & 0.52                                 & 0.6672                           & 17.98                          & 35.42                           & 0.9311                           & 1.00                                 & 0.5682                           & 29.80                           \\
\textbf{Proposed} &                           & \textbf{39.74} & \textbf{0.9781} & \textbf{0.41}       & \textbf{0.6500}   & \textbf{12.17} & \textbf{43.69} & \textbf{0.9923} & \textbf{0.27}       & \textbf{0.7750}  & \textbf{5.54} & \textbf{41.71} & \textbf{0.9852} & \textbf{0.34}       & \textbf{0.7125} & \textbf{8.86}  \\ \hline
CAE                                & \multirow{7}{*}{75}       & 31.52                           & 0.8938                           & 1.06                                 & 0.3397                           & 86.04                           & 32.08                           & 0.9105                           & 0.97                                 & 0.3767                           & 79.44                          & 31.8                            & 0.9022                           & 1.01                                 & 0.3582                           & 82.74                           \\
ResCNN                             &                           & 20.7                            & 0.2860                           & 12.03                                & 0.2087                           & 558.26                          & 28.89                           & 0.8526                           & 2.54                                 & 0.2528                           & 131.27                         & 24.8                            & 0.5693                           & 7.29                                 & 0.2308                           & 344.76                          \\
DnCNN                              &                           & 21.28                           & 0.3096                           & 10.82                                & 0.2189                           & 487.45                          & 28.87                           & 0.8486                           & 2.44                                 & 0.2812                           & 122.6                          & 25.07                           & 0.5791                           & 6.63                                 & 0.2501                           & 305.02                          \\
MMD                &                            & 26.05                           & 0.7165                           & 3.34                                 & 0.4733                           & 412.95                          & 26.05                           & 0.7165                           & 3.34                                 & 0.4733                           & 412.95                         & 26.05                           & 0.7165                           & 3.34                                 & 0.4733                           & 412.95                          \\
DAE                &                             & 21.29                           & 0.3065                           & 11.01                                & 0.2202                           & 486.13                          & 27.40                           & 0.8447                           & 2.61                                 & 0.2427                           & 157.53                         & 24.35                           & 0.5756                           & 6.81                                 & 0.2314                           & 321.83                          \\
DRAN                               &                           & 31.75                           & 0.8151                           & 1.85                                 & 0.4365                           & 51.21                           & 37.31                           & 0.9816                           & 0.56                                 & 0.6492                           & 20.00                             & 34.53                           & 0.8984                           & 1.21                                 & 0.5429                           & 35.60                           \\
\textbf{Proposed} &                           & \textbf{38.91} & \textbf{0.9754} & \textbf{0.46}       & \textbf{0.6302} & \textbf{15.02} & \textbf{43.09} & \textbf{0.9918} & \textbf{0.28}       & \textbf{0.7651} & \textbf{6.28} & \textbf{41.00} & \textbf{0.9836} & \textbf{0.37}       & \textbf{0.6977} & \textbf{10.65} \\ \hline
\end{tabular}}
\caption{Quantitative comparison for microscopic image denoising. The proposed method outperforms existing methods on microscopic images as well.}
\label{micTab}
\end{table*}

\subsection{Real-world Medical Image Denoising}
\label{real-ct-denoising}

\begin{figure}[!htb]
\centering
\includegraphics[width=0.7\linewidth,keepaspectratio]{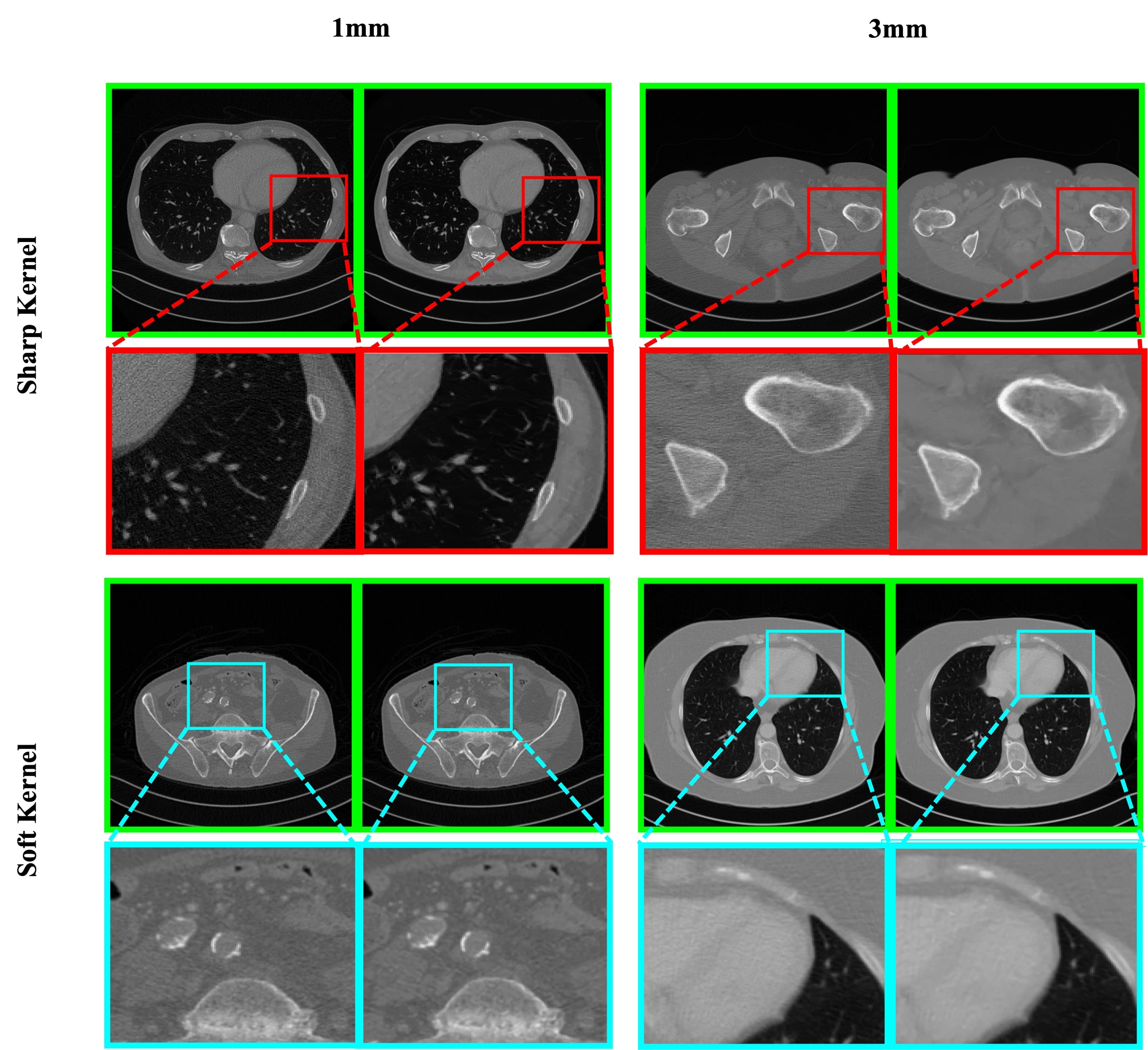}
\caption{  Real-world low-dose CT image denoising with the proposed method. The proposed method can remove real-world noise without losing salient information. In each pair, left: low-dose noisy CT image and right: denoised image. }
\label{realCT}
\end{figure}

Apart from the synthesized noisy images, we evaluated our method with real-world noisy CT images \cite{mccollough2016tu}. Therefore, the model is tuned to real-world noisy medical images. The proposed method has retrained leveraging transfer learning with low-dose sharp kernel CT images \cite{lei2023ct}. The trained model was evaluated with noisy CT images reconstructed with soft and sharp kernels, as shown in Fig. \ref{realCT}. 

The visual results clarify that the proposed method can handle real-world noisy medical images. The proposed method can substantially reduce the low-dose CT noises without showing visual artifacts. In complex spatial regions, it also maintains salient information intact. Overall, the proposed method generates cleaner images and ensures plausible perceptual quality.

\subsection{Ablation Study}

The efficiency of the proposed components has been confirmed with relevant ablation experiments. We evaluated the proposed components with objective and subjective evaluation, similar to the state-of-the-art comparisons. For better readability, we combined all image modalities into a unified dataset. The evaluation metrics, noise levels, and noise patterns remain unchanged. Table. \ref{abl} illustrates the feasibility of the proposed network components. NEN denotes a Noise estimation network (stage I) and TWN denotes the proposed two-stage network without NOB.

\begin{figure}[!htb]
\centering
\includegraphics[width=.8\linewidth,keepaspectratio]{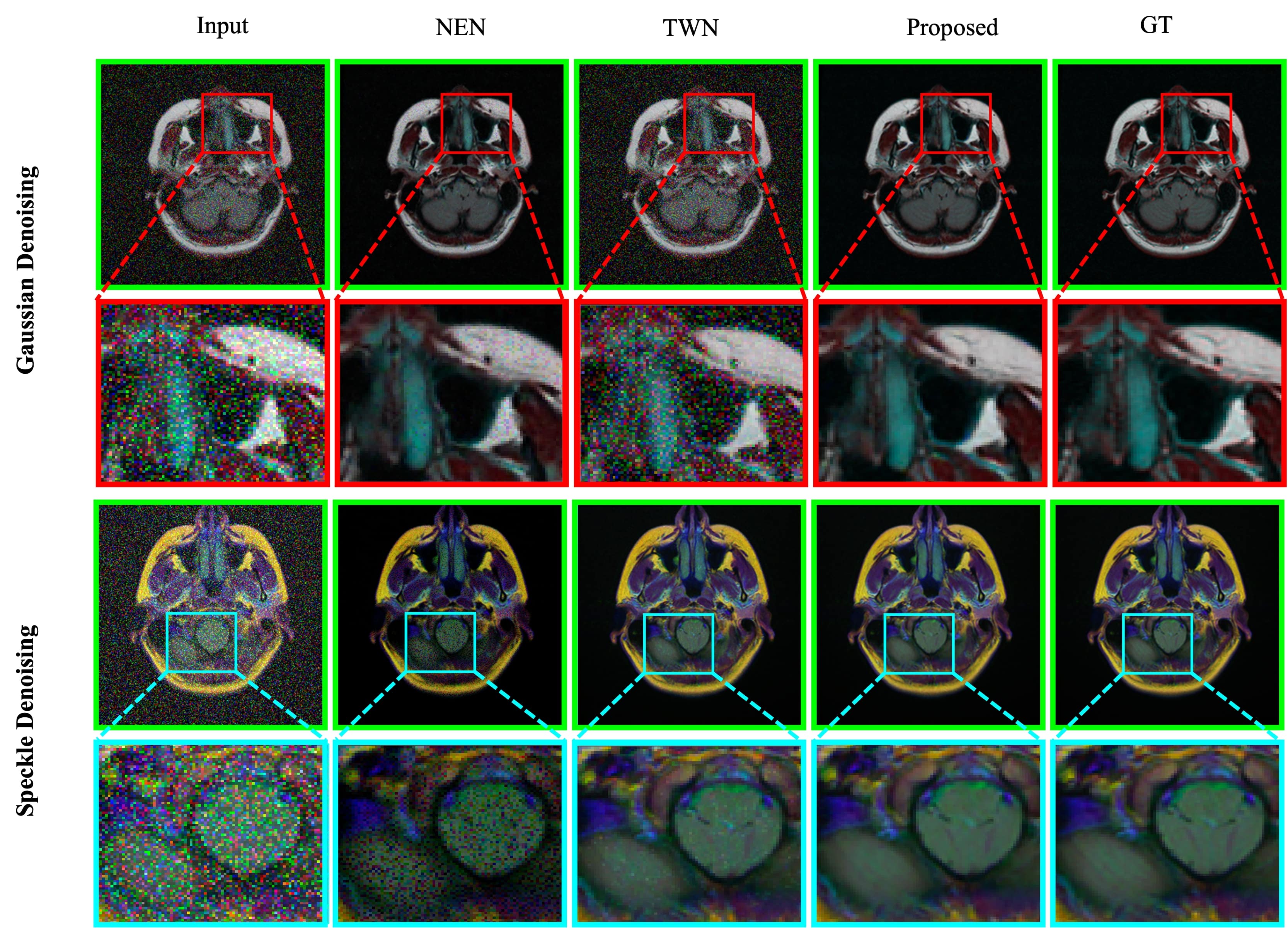}
\caption{  Qualitative evaluation of the proposed method. The top example depicts Gaussian denoising; the bottom illustrates speckle denoising. In each row, Left to right: Noisy Input, the result obtained by NEN (stage I), TWN (two-stage network without NOB), Proposed (two-stage network with NOB), and ground-truth image. It can be seen that the proposed two-stage learning strategy and noise attention help to produce clean images without showing artifacts.}
\label{ablia}
\end{figure}

It is noticeable that the proposed two-stage denoising strategy with self-guided noise attention allows us to achieve state-of-the-art performance for Gaussian and speckle denoising.  Fig. \ref{ablia} also confirms the efficiency of the proposed method in denoising. 

\begin{table*}[!htb]
\centering
\scalebox{.52}{\begin{tabular}{lllllllllllllllll}
\hline
\multirow{2}{*}{\textbf{\textbf{Model}} } & \multirow{2}{*}{\textbf{\textbf{$\sigma$}} } & \multicolumn{5}{l}{\textbf{\textbf{Gaussian}} }                                                & \textbf{\textbf{Speckle}}         &                 &                 &                 &               &  \textbf{\textbf{Average}}         &                 &                 &                 &               \\ \cmidrule(){3-17}
                                &                                 & \textbf{PSNR}  & \textbf{SSIM}   & \textbf{$\Delta E$} & \textbf{VIFP}   & \textbf{MSE}  & \textbf{PSNR}  & \textbf{SSIM}   & \textbf{$\Delta E$} & \textbf{VIFP}   & \textbf{MSE}  & \textbf{PSNR}  & \textbf{SSIM}   & \textbf{$\Delta E$} & \textbf{VIFP}   & \textbf{MSE}  \\ \hline
NEN                         & \multirow{3}{*}{\textbf{10}}    & 36.84          & 0.9508          & 1.51            & 0.6572          & 22.87         & 36.26          & 0.9398          & 1.56            & 0.6421          & 18.83         & 36.55          & 0.9453          & 1.54            & 0.6497          & 20.85         \\
TWN                             &                                 & 34.38          & 0.8690          & 2.47            & 0.5605          & 26.18         & 39.38          & 0.9749          & 0.94            & 0.7338          & 11.18         & 36.88          & 0.9220          & 1.70            & 0.6471          & 18.68         \\
\textbf{Proposed}                    &                                 & \textbf{41.51} & \textbf{0.9811} & \textbf{0.69}   & \textbf{0.7374} & \textbf{6.57} & \textbf{43.58} & \textbf{0.9853} & \textbf{0.54}   & \textbf{0.7936} & \textbf{5.50} & \textbf{42.54} & \textbf{0.9832} & \textbf{0.61}   & \textbf{0.7655} & \textbf{6.04} \\ \hline
Stage-I                         & \multirow{3}{*}{\textbf{25}}    & 36.71          & 0.9454          & 1.47            & 0.6437          & 23.58         & 33.68          & 0.8999          & 2.21            & 0.5623          & 40.36         & 35.19          & 0.9226          & 1.84            & 0.6030          & 31.97         \\
WNN                             &                                 & 31.64          & 0.7892          & 3.60            & 0.4842          & 61.63         & 39.92          & 0.9752          & 0.93            & 0.7337          & 10.29         & 35.78          & 0.8822          & 2.27            & 0.6089          & 35.96         \\
\textbf{Proposed}                    &                                 & \textbf{41.08} & \textbf{0.9799} & \textbf{0.71}   & \textbf{0.7250} & \textbf{7.17} & \textbf{43.03} & \textbf{0.9846} & \textbf{0.56}   & \textbf{0.7841} & \textbf{5.73} & \textbf{42.05} & \textbf{0.9823} & \textbf{0.63}   & \textbf{0.7546} & \textbf{6.45} \\ \hline
Stage-I                         & \multirow{3}{*}{\textbf{50}}    & 35.26          & 0.9153          & 1.77            & 0.6072          & 41.72         & 31.07          & 0.8315          & 3.21            & 0.4807          & 102.14        & 33.16          & 0.8734          & 2.49            & 0.5440          & 71.93         \\
TWN                             &                                 & 28.88          & 0.7034          & 5.03            & 0.4189          & 155.72        & 39.57          & 0.9731          & 1.00            & 0.7203          & 11.08         & 34.22          & 0.8382          & 3.01            & 0.5696          & 83.40         \\
\textbf{Proposed}                    &                                 & \textbf{40.71} & \textbf{0.9789} & \textbf{0.73}   & \textbf{0.7160} & \textbf{7.84} & \textbf{42.59} & \textbf{0.9840} & \textbf{0.59}   & \textbf{0.7754} & \textbf{6.03} & \textbf{41.65} & \textbf{0.9814} & \textbf{0.66}   & \textbf{0.7457} & \textbf{6.94} \\ \hline
Stage-I                         & \multirow{3}{*}{\textbf{75}}    & 33.85          & 0.8819          & 2.22            & 0.5676          & 69.43         & 29.25          & 0.7796          & 4.02            & 0.4262          & 172.64        & 31.55          & 0.8308          & 3.12            & 0.4969          & 121.04        \\
TWN                             &                                 & 26.86          & 0.6430          & 6.27            & 0.3747          & 277.65        & 38.74          & 0.9656          & 1.18            & 0.6961          & 14.27         & 32.80          & 0.8043          & 3.72            & 0.5354          & 145.96        \\
\textbf{Proposed}                    &                                 & \textbf{40.42} & \textbf{0.9780} & \textbf{0.74}   & \textbf{0.7092} & \textbf{8.44} & \textbf{42.25} & \textbf{0.9835} & \textbf{0.61}   & \textbf{0.7684} & \textbf{6.34} & \textbf{41.33} & \textbf{0.9807} & \textbf{0.68}   & \textbf{0.7388} & \textbf{7.39} \\ \hline
\end{tabular}}
\caption{Quantitative evaluation of the proposed two-stage method. The novel components help to achieve state-of-the-art performance for multi-modal medical image denoising. }
\label{abl}
\end{table*}

\begin{figure*}[!htb]
\centering
\includegraphics[width=0.95\linewidth,keepaspectratio]{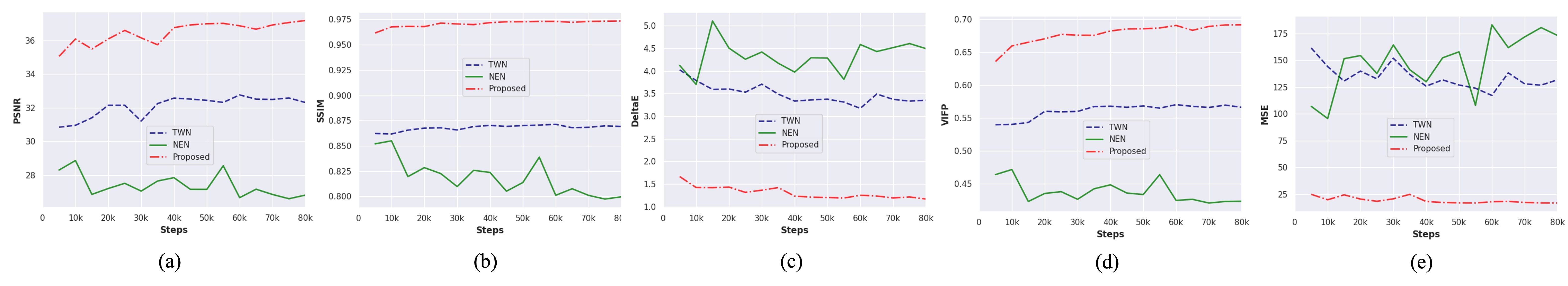}
\caption{ Validation results of the proposed method and its variants. (a) PSNR vs. steps. (b) SSIM vs. steps. (C) $\Delta E$ vs. steps. (d) VIFP vs. steps. (e) MSE vs. steps. }
\label{val_graph}
\end{figure*}

In addition to the ablation experiments, we visualized the validation performance of the proposed network over the training period. Fig. \ref{val_graph} illustrates the validation performance of the proposed method and its variants for speckle and Gaussian denoising.  It is perceptible that the proposed components help the proposed method with two-stage denoising and noise attention help to achieve a stable training result with improved denoising performance in all evaluation metrics. 

\subsection{Inference Analysis}
Despite showing a significant performance gain over state-of-the-art denoising methods, the proposed two-stage denoising method is computationally efficient. Table. \ref{inference} illustrates the proposed method comprises only 586,242 trainable parameters with 38.34 G Macs operations. In our setup, it takes only 25.73 ms to denoise an input image with $256 \times 256 \times 3$ pixels. Overall, it reveals the feasibility of the proposed method for real-world lightweight applications \cite{mujtaba2022client,kim2020compression}. 

\begin{table}[!htb]
\centering
\scalebox{.6}{\begin{tabular}{llll}
\hline
\textbf{\textbf{Model}}                 & \textbf{NEN} & \textbf{RN} & \textbf{Two-stage} \\ \hline
\textbf{Trainable Parameters} & 283,776          & 302,466           & 586,242          \\
\textbf{MAC(G)}               & 18.56            & 19.78             & 38.34            \\
\textbf{Memory Consumption (MB)}   & 835.33           & 900.90            & 1,736            \\
\textbf{Inference Time (ms)}       & 9.96             & 15.77             & 25.73           \\ \hline
\end{tabular}}
\caption{Trainable parameters and inference time of the proposed method. Despite showing significant performance gain, the proposed method is also computationally lightweight. }
\label{inference}
\end{table}

\subsection{Discussion}

A generic method for denoising in medical imaging offers several advantages. This simplified approach can effectively utilize cross-modal domain knowledge to enhance denoising performance. This study incorporates a novel noise simulation method based on real-world scenarios. However, in extreme cases, the proposed method can illustrate shortcomings due to the lack of real-world data samples. The limitation of the proposed method indicates the future directions for multi-modal medical image-denoising. A real-world dataset could help future studies develop a robust real-world generic denoising solution. The modal learning ability of the proposed method also discloses the possibility of learning medical image denoising by leveraging federated learning. It could allow the process of learning denoising in a decentralized manner to reduce data dependency. 

\section{Conclusion}
\label{conclusion}
 This study proposed a two-stage deep denoising method for learning multi-pattern noise from multi-modal medical images. The proposed network estimates the residual noise in Stage I and utilizes it in a later stage for refining denoising output by leveraging a novel noise attention mechanism. Such a learning strategy allowed the proposed method to achieve state-of-the-art performance on Gaussian and speckle denoising in numerous medical modalities. In a qualitative comparison, the proposed two-stage network outperforms existing denoising methods and illustrated a performance gain of 7.64 in PSNR, 0.1021 in SSIM, 0.80  in $\Delta E$, 0.1855  in VIFP, and 18.54 in MSE metrics. There are plans to collect a real-world noisy medical dataset to study the proposed method in the foreseeable future.

\section*{Acknowledgement}
All authors declare that they have no known conflicts of interest in terms of competing financial interests or personal relationships that could have an influence or are relevant to the work reported in this paper. This work was supported by a National Research Foundation (NRF) grant funded by the Ministry of Science and ICT (MSIT), South Korea, through the Development Research Program NRF2021R1A2C1014432 and NRF2022R1G1A1010226, and partially supported by Opt-AI Inc., South Korea.

%

\bibliographystyle{IEEEtran}

\end{document}